\documentclass[11pt]{article}

\usepackage{bbold}
\usepackage{amsmath,amssymb,amsbsy,amstext,amsfonts,graphicx,tcolorbox}
\usepackage{microtype}
\usepackage[bookmarksopen=true]{hyperref}
\usepackage{cleveref}
\usepackage{color}
\usepackage[margin=1.5in]{geometry}
\usepackage[T1]{fontenc}
\usepackage{wasysym}
\usepackage[sort&compress,numbers]{natbib}
\usepackage{epigraph}

\newcommand{\Mp}{M_\mathrm{Pl}}

\newcommand{\dd}{\mathrm{d}}
\newcommand{\sdg}{\sqrt{-g}}

\newcommand{\mn}{{\mu\nu}}
\newcommand{\ab}{{ab}}
\newcommand{\AB}{{AB}}

\newcommand{\mL}{\mathcal{L}}
\newcommand{\rs}{r_\mathrm{s}}
\newcommand{\rst}{r_\star}
\newcommand{\jrs}{J^{r_\star}}
\newcommand{\cc}{{\mathrm{c.c.}}}

\newcommand{\barm}{{\bar{m}}}
\newcommand{\Mtwo}{{\mathcal{M}_2}}

\DeclareMathOperator{\tho}{\text{\TH}}

\definecolor{darkblue}{rgb}{0.15,0.35,0.55}
\definecolor{reddish}{rgb}{0.65, 0.2, 0.2}
\definecolor{darkgreen}{RGB}{50,150,0}
\definecolor{greyish2}{rgb}{.96,.96,.96}
\definecolor{calmingskyblue}{RGB}{0,150,255}
\hypersetup{
colorlinks=true,
citecolor=darkblue,
linkcolor=darkblue,
urlcolor=darkblue,
pdfauthor={Adam R. Solomon},
pdftitle={Off-shell duality invariance of Schwarzschild perturbation theory},
pdfsubject={}
}

\makeatletter
\DeclareRobustCommand{\rcite}[1]{%
  \rcite@aux#1,\@nil{#1}%
}
\def\rcite@aux#1,#2\@nil#3{%
  \if\relax#2\relax
    Ref.~\cite{#3}%
  \else
    Refs.~\cite{#3}%
  \fi
}
\makeatother

\numberwithin{equation}{section}

\epigraphnoindent
\begin{document}

\title{Off-shell duality invariance of \\Schwarzschild perturbation theory}
\author{Adam R. Solomon\thanks{\href{mailto:soloma2@mcmaster.ca}{soloma2@mcmaster.ca}}\\\\ \small Department of Physics and Astronomy, McMaster University,\\\small1280 Main Street West, Hamilton ON, Canada\\\\\small Perimeter Institute for Theoretical Physics,\\\small 31 Caroline Street North, Waterloo ON, Canada}

\date{\today}

\maketitle

\begin{abstract}
We explore the duality invariance of the Maxwell and linearized Einstein-Hilbert actions on a non-rotating black hole background. On shell these symmetries are electric-magnetic duality and Chandrasekhar duality, respectively. Off shell they lead to conserved quantities; we demonstrate that one of the consequences of these conservation laws is that even- and odd-parity metric perturbations have equal Love numbers. Along the way we derive an action principle for the Fackerell-Ipser equation and Teukolsky-Starobinsky identities in electromagnetism.
\end{abstract}

\setcounter{tocdepth}{2}
\tableofcontents

\section{Introduction}

\epigraph{The black holes of nature are the most perfect macroscopic objects there are in the universe: the only elements in their construction are our concepts of space and time.}{\textsc{Chandrasekhar} \cite{Chandrasekhar:1985kt}}

The advent in the past decade of gravitational-wave astronomy and black hole imaging have spurred a renewed observational interest in the foundational and endlessly fascinating black hole solutions of general relativity (GR). The Schwarzschild metric describing non-rotating black holes is in a sense gravity's analog of the hydrogen atom in quantum mechanics: it was the first exact solution of Einstein's equations to be discovered,\footnote{The history is remarkable. Einstein published his field equations and an approximate solution accounting for Mercury's observed perihelion advance in November 1915. Schwarzschild read this work while serving on the Russian front, and by December 1915 had obtained his exact solution. Half a year later he died of an autoimmune disease acquired at the front.} and is still often the first solution taught to students of GR.

The humble Schwarzschild metric is, of course, far from sufficient for modelling gravitational-wave events: astrophysical black holes rotate and so are more accurately described by the significantly more complicated Kerr metric, and the two-body problem in general relativity is highly non-linear and requires numerical techniques to solve near the merger. But some progress can be made analytically, particularly during the inspiral and ringdown phases, through a variety of perturbative schemes. Among the simplest is \emph{black hole perturbation theory}, in which the metric is a small perturbation around a black hole background, analogous to the flat-space perturbation theory which is itself an essential topic in introductory GR courses.

Black hole perturbation theory, in other words, is a fundamental problem in GR with significant relevance to modern experiments. In this paper we explore some of the symmetries of this theory, particularly the \emph{Chandrasekhar duality} between even- and odd-parity modes (which arrive to Earth as $+$ and $\times$ polarizations), which most famously manifests itself in the fact that the quasinormal mode spectra of both sectors are identical \cite{Chandrasekhar:1985kt,Berti:2009kk}.

We will have a particular emphasis on symmetries which hold \emph{off shell}, that is, symmetries of the action rather than just of the equations of motion. Our principal motivation for this is the role played by the action in Noether's theorem; it is also relevant for the quantum theory, e.g., \cite{Agullo:2016lkj,Agullo:2018nfv,Kallosh:2021ors,Kallosh:2021uxa}.
For linear theories, which we consider in this work, it is always possible to construct an action from the equations of motion, so the distinction between on- and off-shell symmetries may seem somewhat artificial. Nevertheless there are interesting differences, as is illustrated by the classical example of electric-magnetic duality in Maxwell's theory.

The electromagnetic field is described by the vector potential $A=A_\mu\dd x^\mu$. In terms of the field strength $F=\dd A$, the field equations in vacuum are\footnote{Here $\star$ is the Hodge star, which in coordinates is $\star F_\mn = \frac12\epsilon_{\mn\alpha\beta}F^{\alpha\beta}$.}
\begin{equation}
\dd\star F=0,\qquad \dd F=0.
\end{equation}
The former is Maxwell's equation, and the latter is the Bianchi identity, which is satisfied for all field configurations since $\dd^2=0$. If we perform a \emph{duality transformation}, by sending $F\to\star F$ and $\star F\to- F$,\footnote{In terms of the electric and magnetic fields this is $(E,B)\to(B,-E)$. Note that $\star^2=-1$ on 2-forms in $3+1$ dimensions.} then the Maxwell equation becomes the Bianchi identity and vice versa, leaving the full set of equations invariant. This is a particular case ($\theta=-\pi/2$) of an $SO(2)$ duality invariance of Maxwell's equations,
\begin{equation}
\begin{pmatrix}
F\\\star F
\end{pmatrix} \to
\begin{pmatrix}
\cos\theta & -\sin\theta \\
\sin\theta & \cos\theta
\end{pmatrix}
\begin{pmatrix}
F\\\star F
\end{pmatrix}.
\end{equation}

Since electric-magnetic duality is a continuous symmetry, Noether's theorem tells us there must be an associated conservation law. To find this, one varies the action under a duality transformation with a spacetime-dependent parameter. However this does \emph{not} mean simply varying the Maxwell action $S=\frac14\int\dd^4x\sdg F_\mn^2$ and setting $\delta F_\mn = \epsilon(x)\star F_\mn$, because $A_\mu$ rather than $F_\mn$ is the dynamical variable which we vary in the action to obtain Maxwell's equations.

The Noether procedure requires us to vary $A$ by a functional $\delta A[A]$ implementing the duality symmetry, but it is impossible to construct a $\delta A[A]$ such that $\dd \delta A = \star F$. If there were, we could take an exterior derivative to find $\dd\star F=0$, i.e., Maxwell's equation for $A$, which is precisely what we do not want to assume.\footnote{Equivalently, note that the Maxwell Lagrangian $E^2-B^2$ naively does not appear to be invariant under rotations of $E$ and $B$, which are indeed a symmetry of the action, but does appear to be invariant under \emph{hyperbolic} $(E,B)$ rotations, which are \emph{not} a genuine symmetry.}
The best we can do is construct a symmetry operator $\delta A[A]$ which is only a duality transformation (in the sense that $\dd\delta A = \star \dd A$) on shell; the full expression contains additional terms which vanish when the Maxwell equations are satisfied \cite{calkin1965invariance,Deser:1976iy,Deser:1981fr}. 

Interestingly the off-shell duality transformation is typically \emph{non-local}. To see this we note that we could flip the roles of the Maxwell equation $\dd\star F=0$ and the Bianchi identity $\dd F=0$ by taking the former to define a potential, $\star F = \dd \tilde A$, and the latter to be the field equation for this ``dual potential'' $\tilde A_\mu$.
This dual potential is precisely the symmetry transformation,
\begin{equation}
\delta A[A] = \tilde A,
\end{equation}
where $\tilde A$ is a solution to the first-order equation $\dd \tilde A = \star \dd A$. Since solving this equation requires integration, in general $\tilde A$ will depend non-locally on $A$. For instance, in a gauge where $\delta A_0=0$, the off-shell duality transformation of $A_i$ is \cite{Deser:1976iy}
\begin{equation}
\delta A^i = \nabla^{-2}(\epsilon^{ijk}\partial_jF_{0k}),
\end{equation}
with $\nabla^{-2}$ the inverse spatial Laplacian.
This is a genuine symmetry of the Maxwell action, which can be used to derive conserved quantities, and which coincides with duality transformations $\delta F = \star F$ on shell, i.e., when the Maxwell equations are satisfied.

The goal of this work is to discuss a similar story for the Chandrasekhar duality in black hole perturbation theory. Along the way we will investigate the dynamics of scalar, electromagnetic, and gravitational fields on the Schwarzschild background in two covariant languages designed to exploit its symmetries, the $2+2$ and Geroch-Held-Penrose (GHP) formalisms. These approaches are complementary: the $2+2$ formulation is more intuitive but specifically adapted to a non-rotating black hole, while GHP generalizes straightforwardly to the full Kerr solution and is in a sense ``more fundamental'' in that it is based on the algebraically-special structure of black hole spacetimes. We will further see that objects arising naturally when studying dynamics in the $2+2$ formulation have simple interpretations in GHP language.

The rest of this paper is organized as follows. In \cref{sec:Sch} we review the Schwarzschild solution and introduce the $2+2$ and GHP formalisms. We study the dynamics of a massless scalar field on Schwarzschild in \cref{sec:scalar}, the electromagnetic field in \cref{sec:EM}, and linearized gravity in \cref{sec:gravity}. In \cref{sec:Chandra} we discuss the off-shell Chandrasekhar duality and in \cref{sec:Love} explore its physical consequences for tidal Love numbers, before concluding in \cref{sec:disc}.

\textbf{Conventions:} We work with vacuum general relativity in $3+1$ spacetime dimensions with metric signature $(-,+,+,+)$ and choose an orientation such that $\epsilon_{0123}=\sdg$. We will use Greek letters $\mu,\nu,\cdots$ for four-dimensional spacetime indices, lower-case Latin letters $a,b,...$ for the $(t,r)$ subspace $\mathcal{M}_2$, and upper-case letters $A,B,...$ for the 2-sphere $S^2$.

\section{Schwarzschild background in the $2+2$ and GHP formalisms}
\label{sec:Sch}

The black hole solutions in vacuum four-dimensional general relativity are highly symmetrical. In this section we will review the Schwarzschild metric, on which we will place various field theories, in two formalisms designed to exploit these symmetries in a coordinate-independent manner. The first is the $2+2$ formalism, which treats objects covariantly on the two-sphere and on the $(t,r)$ plane. The second is the GHP formalism, which takes advantage of the algebraically-special (type D) structure of black hole spacetimes in general relativity.

The Schwarzschild metric in Boyer-Lindquist (or Schwarzschild) coordinates is
\begin{equation}
g_\mn\dd x^\mu\dd x^\nu = -f(r) \dd  t^2 + \frac{1}{f(r)}\dd  r^2+r^2 \underbrace{\left(\dd\theta^2+\sin^2\theta\dd\phi^2\right)}_{\dd \Omega_{S^2}^2},\qquad f(r) \equiv 1-\frac\rs r
\label{eq:sphericalmetric}
\end{equation}
with $\rs=2GM$ the Schwarzschild radius and $\dd \Omega^2_{S^2}$ the line element on the unit 2-sphere. As we will see, kinetic terms for fields on a Schwarzschild background are often more conveniently phrased in terms of a ``tortoise coordinate'' $\rst$ defined by
\begin{equation}\label{eq:tortoise}
\dd\rst=\frac{\dd r}{f(r)}.
\end{equation}
The horizon $r=\rs$ is located at $\rst=-\infty$ and spatial infinity $r=\infty$ at $\rst=\infty$.

\subsection{$2+2$ decomposition}

The Schwarzschild spacetime factorizes naturally into two submanifolds: the $(t,r)$ plane $\mathcal{M}_2$ and the 2-sphere $S^2$. This is the basis of the $2+2$ decomposition \cite{Sarbach:2001qq,Martel:2005ir,Chaverra:2012bh}.
Let us write the four-dimensional coordinates as $x^\mu = (x^a,\theta^A)$, where lower-case Latin letters $a,b,...$ run over $(t,r)$ and upper-case letters $A,B,...$ run over $(\theta,\phi)$. The metric factorizes into
\begin{equation}
g_\mn\dd x^\mu\dd x^\nu = g_{ab}\dd x^a \dd x^b + r^2\Omega_{AB}\dd\theta^A\dd\theta^B, \label{eq:2+2}
\end{equation}
with
\begin{equation}
g_{ab}=\begin{pmatrix}
-f&0\\
0&\frac1f
\end{pmatrix},\qquad\Omega_{AB}=\begin{pmatrix}1&0\\0&\sin^2\theta\end{pmatrix}.
\end{equation}
To avoid a clutter of notation, we will use $\nabla_\mu$, $\nabla_a$, and $D_A$ for the covariant derivatives with respect to $g_\mn$, $g_\ab$, and $\Omega_\AB$, respectively, and raise and lower indices with these metrics. We also use the same symbol for $g_\mn$ and $g_\ab$; which one is meant should be clear from context.\footnote{In particular, $\sdg$ represents the square root of the determinant of $g_\mn$ in $\int\dd^4x\sdg$ and of $g_\ab$ in $\int\dd^2x\sdg$.}

The $r$ appearing in \cref{eq:2+2} is a spacetime scalar on $\Mtwo$ and need not be aligned with one of the coordinate directions, though it is in Boyer-Lindquist coordinates. It and the 2-metric $g_\ab$ obey the background Einstein equations,
\begin{equation}
r R = 2 \Box r,\qquad \nabla^a(r\nabla_ar) =r \Box r+(\partial r)^2=1,\qquad \nabla_a\nabla_br=\frac12\Box r  g_\ab, \label{eq:back-eom}
\end{equation}
where $\Box=g^\ab\nabla_a\nabla_b$ and $(\partial r)^2=g^\ab\partial_ar\partial_br$. In coordinates, the Ricci scalar and the norm of $\partial_ar$ are
\begin{equation}
R = \frac{2\rs}{r^3},\qquad (\partial r)^2 = f.\label{eq:M2scalars}
\end{equation}
Note in particular that the latter of these allows us to use $f(r)$ in coordinate-invariant expressions. We will find it convenient at times to use the shorthand
\begin{equation}
r_a = \partial_ar.
\end{equation}

As a consequence of its high degree of symmetry, equations of motion on the Schwarzschild background admit fully separable solutions \cite{Frolov:2017kze}. For a field of integer spin $s$, the general solution for the field variable or an observable constructed from it can be written in the schematic form (e.g., omitting indices)
\begin{equation}
\phi(x^\mu) = \displaystyle\sum_{\ell=|s|}^\infty\displaystyle\sum_{m=-\ell}^\ell \underbrace{ \int \dd\omega e^{-i\omega t} R_{\ell \omega}(r)}_{\phi_{\ell m}(x^a)}\underbrace{\vphantom{\int}\Theta_{\ell m}(\theta)e^{im\phi}}_{S_{\ell m}(\theta^A)}.\label{eq:harmonic-decomp}
\end{equation}
A further consequence of symmetry is that the radial and angular functions $R_{\ell\omega}(r)$ and $\Theta_{\ell m}(\theta)$ obey remarkably similar equations. The main difference is that the periodic boundary conditions on the angular coordinates constrain $S_{\ell m}(\theta,\phi)$ to the class of spherical harmonic functions, which are eigenfunctions of the Laplacian on $S^2$, while $R_{\ell\omega}(r)$ obeys a Schr\"odinger-like equation (typically in terms of the tortoise coordinate $\rst$ rather than $r$).

The spherical harmonics can be categorized by their transformation properties under rotations. In four dimensions, there are two such classes: scalars and vectors.\footnote{Degrees of freedom transforming under the tensor representation are non-dynamical in $D=4$ but are present in higher dimensions.} The scalar harmonics are the familiar spherical harmonics,
\begin{equation}
S_{\ell m} = Y_{\ell m}(\theta,\phi) \propto P^m_\ell(\cos\theta)e^{im\phi},
\end{equation}
with $P^m_\ell(x)$ the associated Legendre polynomials. The vector harmonics decompose into longitudinal and transverse, or electric and magnetic, pieces, which are related to the scalar harmonics by
\begin{subequations}
\begin{align}
E_{A,\ell m} &= D_A Y_{\ell m},\\
B_{A,\ell m} &= -\epsilon_{AB}D^B Y_{\ell m},\label{eq:BA}
\end{align}
\end{subequations}
with $\epsilon_{AB}$ the Levi-Civita tensor on the 2-sphere, $\epsilon_{\theta\phi}=\sin\theta$. In coordinates these are
\begin{subequations}
\begin{align}
E_{A,\ell m}\dd \theta^A &= \partial_\theta Y_{\ell m}\dd\theta+\partial_\phi Y_{\ell m}\dd\phi,\\
B_{A,\ell m}\dd\theta^A &= -\csc\theta\partial_\phi Y_{\ell m} \dd\theta + \sin\theta\partial_\theta Y_{\ell m}\dd\phi.
\end{align}
\end{subequations}

The scalar harmonics obey the Laplace equation on the 2-sphere with eigenvalue $-\ell(\ell+1)$,
\begin{align}
D^2Y_{\ell m} &= \frac{1}{\sqrt\Omega}\partial_A\left(\sqrt\Omega \Omega^{AB}\partial_BY_{\ell m}\right)\nonumber\\
& = -\ell(\ell+1)Y_{\ell m},
\end{align}
where $\Omega\equiv\det(\Omega_\AB)=\sin^2\theta$, while the vector harmonics $V_A = (E_A,B_A)$ are eigenfunctions with eigenvalue $1-\ell(\ell+1)$,
\begin{equation}
D^2 V_A^{\ell m} = -\left[\ell(\ell+1)-1\right]V_A^{\ell m}.
\end{equation}
The spacetime integration measure appearing in a four-dimensional action contains the 2-sphere integration measure $\dd\Omega$,\footnote{We remind the reader that in our notation, $\int\dd^4x\sdg = \int\dd^4x\sqrt{-\det g_\mn}$ while $\int\dd^2x\sdg = \int\dd^2x\sqrt{-\det g_\ab}$.}
\begin{equation}
\int\dd^4x\sdg = \int\dd^2x\sdg r^2\dd\Omega,\qquad \int\dd\Omega\equiv\int_{\theta=0}^\pi\int_{\phi=0}^{2\pi}\sin\theta\dd\theta\dd\phi.
\end{equation}
We will be able to integrate over $S^2$ in actions on Schwarzschild using the orthonormality relations of the spherical harmonics,
\begin{subequations}
\begin{align}
\int\dd\Omega Y_{\ell m}Y_{\ell'm'} &= \delta_{\ell\ell'}\delta_{mm'}, \\
\int\dd\Omega V_{A,\ell m}V^A_{\ell'm'} &= \ell(\ell+1)\delta_{\ell\ell'}\delta_{mm'},\\
\int\dd\Omega E_{A,\ell m}B^A_{\ell'm'} &= 0.
\end{align}
\end{subequations}

\subsection{Geroch-Held-Penrose (GHP) formalism}

In this subsection we describe an alternative formalism for leveraging the symmetry of black hole backgrounds: the Geroch-Held-Penrose (GHP) formalism, which is itself built on the famous Newman-Penrose (NP) approach. While this approach is somewhat more arcane than the $2+2$ formalism,\footnote{Due at least in part to its heavy use of Icelandic runes.} it more directly makes use of the fundamental property underpinning the ``magic'' of the Schwarzschild and Kerr spacetimes, namely the fact that they are \emph{algebraically special}.

\subsubsection{Newman-Penrose}

Recall that the Weyl tensor $C_{\mn\alpha\beta}$ of a generic spacetime has four principal null directions;\footnote{Principal null directions are null vectors $l^\mu$ satisfying $l^\nu l_{[\rho}C_{\mu]\nu\alpha[\beta}l_{\sigma]}l^\alpha = 0$ \cite{Newman:2009}.} algebraically-special spacetimes are those where one or more of the four are degenerate. The Kerr black hole is of algebraic \emph{type D},
with two singly-degenerate principal null directions. These special vectors, $l^\mu$ and $n^\mu$, point along outgoing and ingoing null rays, respectively. In the Schwarzschild case they live on $\Mtwo$,
\begin{equation}
l_\mu\dd x^\mu = l_a\dd x^a, \quad n_\mu\dd x^\mu = n_a\dd x^a,
\end{equation}
and in fact can be thought of as zweibeins for the 2-metric,
\begin{equation}
g_\ab = -l_an_b-n_al_b.
\end{equation}
To complete the picture, we include null vectors parametrizing $S^2$: a complex vector $m^\mu$ and its complex conjugate $\barm^\mu$, with $m_\mu \dd x^\mu = m_A\dd\theta^A$. These four vectors together comprise a complex null tetrad $e^\mathbf{a}_\mu = (l_\mu,n_\mu,m_\mu,\bar m_\mu)$, in the sense that\footnote{This is the usual vielbein relation $g_\mn = \eta_\mathbf\ab e^\mathbf a_\mu e^\mathbf b_\nu$ with the internal Minkowski metric written in the form
\[ \eta_\mathbf\ab = \begin{pmatrix}
0&-1&0&0\\
-1&0&0&0\\
0&0&0&1\\
0&0&1&0
\end{pmatrix}. \]
Here bold lowercase Latin letters represent 4D internal Lorentz indices.}
\begin{equation}
g_\mn = -2l_{(\mu}n_{\nu)} + 2m_{(\mu}\barm_{\nu)}\label{eq:metric-tetrad}.
\end{equation}
The vielbeins are normalized so that all of their inner products vanish except for
\begin{equation}
l_\mu n^\mu = -1,\quad m_\mu \barm^\mu = 1.
\end{equation}

This setup does not completely fix $(l^\mu,n^\mu,m^\mu,\barm^\mu)$, as there is some residual Lorentz invariance. Insisting that $\ell^\mu$ and $n^\mu$ remain principal null directions leaves a two-parameter symmetry comprising boosts of $l$ and $n$,
\begin{equation}
l^\mu \to \alpha l^\mu,\quad n^\mu \to \alpha^{-1}n^\mu, \label{eq:tetrad-tr-rescale}
\end{equation}
and rotations of $m$ and $\barm$,
\begin{equation}
m^\mu\to e^{i\beta}m^\mu,\quad\barm^\mu\to e^{i\beta}\barm^\mu,\label{eq:tetrad-S2-rescale}
\end{equation}
with $\alpha$ and $\beta$ real functions. We will choose the \emph{Carter tetrad} \cite{Carter:1987hk},
\begin{subequations}
\begin{align}
l_\mu\dd x^\mu &= \frac{1}{\sqrt2}\left(-\sqrt f \dd t + \frac{1}{\sqrt f}\dd r\right),\\
n_\mu\dd x^\mu &= \frac{1}{\sqrt2}\left(-\sqrt f \dd t - \frac{1}{\sqrt f}\dd r\right),\\
m_\mu\dd x^\mu &= \frac{r}{\sqrt2}\left(\dd\theta+i\sin\theta\dd\phi\right),\\
\barm_\mu\dd x^\mu &= \frac{r}{\sqrt2}\left(\dd\theta-i\sin\theta\dd\phi\right).
\end{align}
\end{subequations}
The frequently-used Kinnersley tetrad \cite{Kinnersley:1969zza} is related by a rescaling \eqref{eq:tetrad-tr-rescale} with $\alpha = \sqrt{f/2}$. The Carter tetrad is particularly useful for our purposes as it maintains symmetries of the background which can be obscured in other bases \cite{Pound2020}.

In the Newman-Penrose formalism one works with spacetime scalars obtained by projection along the null directions. For instance the Weyl tensor $C_{\mn\alpha\beta}$ is efficiently encoded in five complex Weyl scalars, which are the ``components'' of the Weyl tensor in the complex null basis,
\begin{align}
\Psi_0= C_{lmlm},\quad\Psi_1=C_{lnlm},\quad\Psi_2=C_{lm\barm n},\quad\Psi_3=C_{ln\barm n},\quad \Psi_4 = C_{n\barm n\barm},\label{eq:weylscalars}
\end{align}
where $C_{lm\barm n}=C_{\mn\alpha\beta}l^\mu m^\nu \barm^\alpha n^\beta$ and so on. (In general we will use the notation $V_\mu l^\mu=V_l$, etc.)
For type-D spacetimes the only non-vanishing Weyl scalar is $\Psi_2$, providing a remarkably compact characterization of the full Riemann tensor. In the Schwarzschild case, the value of $\Psi_2$ in coordinates is\footnote{The resemblance to the Ricci scalar on $\Mtwo$, cf. \cref{eq:M2scalars}, is not accidental. Using $R_{aAbB} = -r \nabla_a\nabla_br \Omega_\AB$ \cite{Chaverra:2012bh}, we find $\Psi_2 = R_{aABb}l^an^bm^A\barm^B = \frac1r\nabla_a\nabla_brl^an^b = -\frac14R$.}
\begin{equation}
\Psi_2 = -\frac{\rs}{2r^3}.
\end{equation}

\subsubsection{Geroch-Held-Penrose}

The GHP formalism soups up the NP formalism by working only with quantities and operators which have simple transformation properties under the residual Lorentz invariance \eqref{eq:tetrad-tr-rescale}--\eqref{eq:tetrad-S2-rescale}. Defining $\lambda^2=\alpha e^{i\beta}$, we will insist on working with tensors $\Phi$ that transform under \cref{eq:tetrad-tr-rescale,eq:tetrad-S2-rescale} as
\begin{equation}
\Phi \to \lambda^p\bar\lambda^q\Phi.
\end{equation}
Such a quantity is said to have \emph{GHP type} $\{p,q\}$. They are also called spin- and/or boost-weighted, where the spin weight is $s=(p-q)/2$ and the boost weight is $b=(p+q)/2$.

The residual Lorentz transformations \eqref{eq:tetrad-tr-rescale}--\eqref{eq:tetrad-S2-rescale} do not exhaust the symmetry in choosing a tetrad, which is invariant under several discrete tetrad interchanges: \emph{complex conjugation}, which swaps $m^\mu$ and $\barm^\mu$; the \emph{prime} ($'$) operation, which interchanges both $l\leftrightarrow n$ and $m\leftrightarrow \barm$; and, less obviously, the \emph{star} ($\star$) operation, $(l,n,m,\barm)\to(m,-\barm,-l,n)$, which we will not use. These discrete invariances allow for a particularly economical description of field equations, since one equation implies its prime, conjugate, and prime conjugate versions.

Scalars with well-defined GHP type include the Weyl scalars, which inherit their GHP types from the various factors of $l^\mu$, etc., in their definitions \eqref{eq:weylscalars},\footnote{Tensors like $C_{\mu\nu\alpha\beta}$ are \textit{a priori} unweighted.} as well as the spin coefficient $\rho$,\footnote{And by extension $\rho'$, $\bar\rho$, and $\bar\rho'$, although for Schwarzschild $\rho$ and $\rho'$ are real.}
\begin{equation}
\rho = -\barm^\mu m^\nu \nabla_\mu l_\nu,
\end{equation}
which is of GHP type $\{1,1\}$. Examples of scalars \emph{without} a well-defined GHP type include the spin coefficients $\beta$ and $\epsilon$ (and their primes and conjugates),
\begin{align}
\beta &= \frac12\left(m^\mu\barm^\nu\nabla_\mu m_\nu - m^\mu n^\nu\nabla_\mu l_\nu\right),\\
\epsilon &= \frac12\left(l^\mu\barm^\nu\nabla_\mu m_\nu - l^\mu n^\nu\nabla_\mu l_\nu\right).
\end{align}
These are the only non-zero spin coefficients for Schwarzschild and completely describe the spin connection. In the Carter tetrad they take the coordinate values
\begin{equation}
\rho = -\rho' = -\frac{\sqrt f}{\sqrt2r},\quad \beta=\beta'=\frac{\cot\theta}{2\sqrt{2}r},\quad \epsilon = -\epsilon' = \frac{\rs}{4\sqrt{2f}r^2}.
\end{equation}

Analogously to the non-coordinate-invariant Christoffel symbols, $\beta$ and $\epsilon$ can be used to construct covariant derivative operators with well-defined GHP type. Unfortunately, the use of Icelandic runes for these operators is firmly embedded in the literature:
\begin{align}
&&\tho &= l^\mu \nabla_\mu - p\epsilon - q\bar\epsilon,&\eth &= m^\mu \nabla_\mu - p\beta + q\bar\beta',&&\\
&& \tho'&= n^\mu\nabla_\mu +p\epsilon'+q\bar\epsilon'  & \eth'&= \barm^\mu\nabla_\mu +p\beta'-q\bar\beta&&\nonumber
\end{align}
The operator $\tho$ sends a GHP type $\{p,q\}$ object to one with type $\{p+1,q+1\}$, $\tho'$ to $\{p-1,q-1\}$, $\eth$ to $\{p+1,q-1\}$, and $\eth'$ to $\{p-1,q+1\}$. Note that $\tho$ and $\tho'$ raise and lower the boost weight, while $\eth$ and $\eth'$ raise and lower the spin weight. For the Carter tetrad in Schwarzschild, the GHP derivatives take the coordinate form \cite{Pound2020}
\begin{align}
&& \tho &= \frac{1}{\sqrt{2f}}\left(\partial_t+f\partial_r-\frac{b\rs}{2r^2}\right),& \eth &= \frac{1}{\sqrt2r}\left(\partial_\theta+i\csc\theta\partial_\phi-s\cot\theta\right) &&\\
&&\tho' &= \frac{1}{\sqrt{2f}}\left(\partial_t-f\partial_r-\frac{b\rs}{2r^2}\right) ,&\eth' &= \frac{1}{\sqrt2r}\left(\partial_\theta-i\csc\theta\partial_\phi+s\cot\theta\right). &&\nonumber
\end{align}
Note also that these derivatives have non-trivial commutators,
\begin{equation}
[\tho,\tho'] = -2b\Psi_2,\quad [\tho,\eth] = \rho\eth,\quad [\eth,\eth'] = 2s(\Psi_2+\rho\rho') = -\frac{s}{r^2},\label{eq:GHPcommutators}
\end{equation}
along with their primes and complex conjugates.

In this language, the scalar spherical harmonics are eigenfunctions of $\eth\eth'$,
\begin{equation}
\eth\eth'Y = -\frac{\ell(\ell+1)}{2r^2}Y,
\end{equation}
and are a special case of the \emph{spin-weighted} spherical harmonics,
\begin{equation}
\frac12\left(\eth'\eth+\eth\eth'\right)Y_s = -\frac{\ell(\ell+1)-s^2}{2r^2}Y_s,
\end{equation}
which can be obtained from the scalar harmonics by raising and lowering the spin weight with $\eth$ and $\eth'$,
\begin{equation}
\eth Y_s = -\frac{\sqrt{\ell(\ell+1)-s(s+1)}}{\sqrt2r}Y_{s+1}, \quad \eth' Y_s = \frac{\sqrt{\ell(\ell+1)-s(s-1)}}{\sqrt2r}Y_{s-1}.
\end{equation}
The $|s|=1$ spin-weighted harmonics are related to the vector harmonics by \cite{Pound2020}
\begin{subequations}
\begin{align}
E_A &= \frac{\sqrt{\ell(\ell+1)}}{2}\left(Y_{-1} \tilde m_A - Y_1 \bar {\tilde m}_A\right),\\
B_A &= -i\frac{\sqrt{\ell(\ell+1)}}{2}\left(Y_{-1}\tilde  m_A + Y_1 \bar {\tilde m}_A\right),
\end{align}
\end{subequations}
where $\tilde m_A\dd\theta^A = \dd\theta+i\sin\theta\dd\phi$.

\section{Massless scalar}
\label{sec:scalar}

We want to compute the action for linearized gravity on Schwarzschild, performing separation of variables and utilizing the $2+2$ decomposition. Many of the basic steps of the computation are present in the simpler cases of a scalar and vector field, so we will work our way up to gravity one integer step in spin at a time.

The action for a massless scalar is\footnote{We remind the reader that to avoid a clutter of notation we are using $\sdg$ for both $\sqrt{-\det g_\mn}$ and $\sqrt{-\det g_\ab}$, with the meaning clear depending whether we are integrating over $\dd^4x$ or $\dd^2x$.
Note also that in Boyer-Lindquist coordinates, $\sqrt{-\det g_\ab} = 1$.}
\begin{equation}
S = -\frac12\int\dd^4x\sdg(\partial_\mu\phi)^2. \label{eq:KG4D}
\end{equation}
The field $\phi$ admits a spherical harmonic expansion of the form \eqref{eq:harmonic-decomp},
\begin{equation}
\phi(x^\mu) = \displaystyle\sum_{\ell=0}^\infty \displaystyle\sum_{m=-\ell}^\ell \phi_{\ell m} (x^a) Y_{\ell m}(\theta^A).
\end{equation}
Inserting this into \cref{eq:KG4D} and integrating over $S^2$ we find a sum over actions for each $(\ell,m)$ mode,
\begin{align}
S &= -\frac12\int\dd^2x\sdg r^2\int\dd\Omega \left[(\partial_a\phi)^2+r^{-2}(\partial_A\phi)^2\right]\nonumber\\
&=-\frac12\int\dd^2x \sdg\displaystyle\sum_{\ell\ell'mm'} \int\dd\Omega\left[r^2\partial_a\phi_{\ell m}\partial^a\phi_{\ell' m'}  +\ell(\ell+1)\phi_{\ell m}\phi_{\ell'm'} \right]Y_{\ell m}Y_{\ell'm'} \nonumber\\
&=-\frac12\int\dd^2x \sdg\displaystyle\sum_{\ell,m}\left[r^2(\partial\phi_{\ell m})^2 +\ell(\ell+1)\phi_{\ell m}^2 \right]\nonumber\\&\equiv \displaystyle\sum_{\ell,m}S_{\ell m}.
\end{align}
To simplify notation, we will drop the $\ell m$ subscripts and focus on an individual mode, with the summation over all modes implied. This is kosher because in linear theories modes of different $(\ell,m)$ decouple.

The $2D$ field $\phi$ is not canonically normalized, as its kinetic term is multiplied by a factor of $r^2$. We can remove this with a field redefinition \cite{Chaverra:2012bh,Hui:2020xxx},
\begin{equation}
\psi\equiv r \phi,
\end{equation}
in terms of which the action is
\begin{equation}
\boxed{S = \int\dd^2x\sqrt{-g}\left[-\frac12(\partial\psi)^2-\frac{1}{2r^2}\left(\ell(\ell+1)+\frac{\rs}{r}\right)\psi^2\right].}
\end{equation}
We identify the usual scalar potential on a Schwarzschild background \cite{Berti:2009kk},
\begin{equation}
V(r) = \frac{\ell(\ell+1)}{r^2}+\frac{\rs}{r^3}.
\end{equation}
If we drop our insistence on covariance and write the action in terms of the coordinates $(t,r)$,
\begin{equation}
S = \int \dd t \dd r \left(\frac12f^{-1}(\partial_t\psi)^2-\frac12f(\partial_r\psi)^2-\frac12V(r)\psi^2\right),
\end{equation}
we find that the kinetic and gradient terms again have nonstandard factors in front. To canonically normalize we transform to the tortoise coordinate $\dd r = f\dd r_\star$ \cite{Hui:2020xxx},
\begin{equation}
S = \int \dd t \dd r_\star \left(\frac12(\partial_t\psi)^2-\frac12(\partial_{r_\star}\psi)^2-\frac12V(r)\psi^2\right).
\end{equation}

For completeness let us write the action \eqref{eq:KG4D} in GHP language. Writing the metric in terms of the null vectors, cf. \cref {eq:metric-tetrad}, we have
\begin{align}
S &= -\frac12\int\dd^4x\sdg g^\mn\partial_\mu\phi\partial_\nu\phi \nonumber\\
&= \int\dd^4x\sdg\left(l^\mu n^\nu - m^\mu \barm^\nu\right)\partial_\mu\phi\partial_\nu\phi \nonumber\\
&= \int \dd^4x\sdg \left(\tho\phi\tho'\phi-\eth\phi\eth'\phi\right).
\end{align}
If we separate variables and integrate over the 2-sphere, then the action for a single mode is
\begin{equation}
S_{\ell m} = \int \dd^2x\sdg \left(r^2\tho\phi\tho'\phi-\frac{\ell(\ell+1)}{2}\phi^2\right).
\end{equation}

\section{Electromagnetism}
\label{sec:EM}

The next step on the road to gravity, which is the spin-2 case, is the spin-1 case, which is electromagnetism. The Maxwell action is
\begin{equation}
S = -\frac14\int\dd^4x\sdg F_\mn^2,\qquad F_\mn = 2\partial_{[\mu}A_{\nu]}.
\end{equation}
The vector potential is a superposition of separable solutions:
\begin{equation}
A^\mu = \displaystyle\sum_{\ell=1}^\infty\displaystyle\sum_{m=-\ell}^\ell A^\mu_{\ell m}.
\end{equation}
Herein we will focus on a single mode and drop $\ell m$ subscripts, with the summation implied. Under a $2+2$ decomposition the vector potential is
\begin{equation}
A_\mu\dd x^\mu = A_a(x^a)Y\dd x^a + a(x^a)B_A\dd\theta^A.
\end{equation}
Here we have used our gauge freedom to remove the longitudinal mode, which is proportional to $E_A\dd\theta^A$.

Gauge invariance adds a wrinkle that was not present for the scalar: in order to avoid losing information when fixing a gauge at the level of the action rather than the equations of motion, one must make a \emph{complete} gauge fixing, in the sense that there are no integration constants left when fixing a gauge vector (rather than necessarily that all gauge freedom is exhausted, although we will insist on this too) \cite{Lagos:2013aua,Motohashi:2016prk}. Our gauge choice satisfies this requirement \cite{Hui:2020xxx}.

Performing separation of variables and integrating over the 2-sphere, we obtain
\begin{equation}
S = \int\dd^2x\sqrt{-g}\mathcal{L},
\end{equation}
where
\begin{equation}
\boxed{\mathcal{L} = \underbrace{\vphantom{\bigg[}-\frac14r^2F_{ab}^2-\frac12\ell(\ell+1)A_a^2\vphantom{\bigg]}}_{\mathcal{L}_\mathrm{even}} \underbrace{\vphantom{\bigg[}-\frac12\ell(\ell+1)\left[(\partial a)^2 + \frac{\ell(\ell+1)}{r^2}a^2\right]\vphantom{\bigg]}}_{\mathcal{L}_\mathrm{odd}} \label{eq:L-EM}.}
\end{equation}
We see that the even-parity (or electric) field $A_a$ and the odd-parity (or magnetic) field $a$ decouple.

The even sector has only one dynamical degree of freedom but depends on two variables $A_a$. To isolate this dynamical field we integrate in an auxiliary variable $\lambda(t,r)$:
\begin{align}
\mathcal{L}_\mathrm{even,aux} &= \mathcal{L}_\mathrm{even} + \frac14r^2\left(F_{ab}+r^{-2}\lambda\epsilon_{ab}\right)^2\nonumber\\
&= \lambda \epsilon^{ab}\partial_a A_b-\frac12\frac{\lambda^2}{r^2}-\frac12\ell(\ell+1)A_a^2.
\end{align}
The $\lambda$ equation of motion fixes it to be proportional to $F_\ab$ on-shell,
\begin{equation}
\lambda = r^2\epsilon^{ab}\partial_a A_b = -r^2 F_{tr}.\label{eq:lambdasol}
\end{equation}
Inserting this back into $\mL_\mathrm{even,aux}$ we obtain $\mL_\mathrm{even}$, establishing their dynamical equivalence. However we can also obtain an action for $\lambda$ alone by integrating out $A_a$ using its equation of motion,
\begin{equation}
A^a = \frac{1}{\ell(\ell+1)}\epsilon^{ab}\partial_b\lambda,\label{eq:Aasol}
\end{equation}
and plugging back into the action,
\begin{equation}
\mathcal{L} = -\frac{1}{2\ell(\ell+1)}(\partial\lambda)^2-\frac12\frac{\lambda^2}{r^2}- \frac12\ell(\ell+1)(\partial a)^2- \frac12\frac{\ell^2(\ell+1)^2}{r^2}a{}^2.
\end{equation}
We canonically normalize the fields by scaling out appropriate factors of $\sqrt{\ell(\ell+1)}$,
\begin{equation}
\psi_+ \equiv \frac{\lambda}{\sqrt{\ell(\ell+1)}},\qquad \psi_-\equiv \sqrt{\ell(\ell+1)}a,
\end{equation}
so that
\begin{equation}
\mathcal{L} = \displaystyle\sum_\pm\left[-\frac12(\partial\psi_\pm)^2-\frac12\frac{\ell(\ell+1)}{r^2}\psi_\pm^2\right].\label{eq:L-EM-aux}
\end{equation}
We conclude that $\psi_\pm$ are the ``master variables'' for the electric ($+$) and magnetic ($-$) sectors (see also \rcite{Hui:2020xxx}), each satisfying a Schr\"odinger equation with the usual vector potential \cite{Berti:2009kk}.

\subsection{Electric-magnetic duality}

The Lagrangian \eqref{eq:L-EM-aux} is manifestly invariant under electric-magnetic duality, which acts as a rotation on the vector $(\psi_+,\psi_-)^T$. The infinitesimal version is
\begin{equation}
\delta (\psi_+,\psi_-) = (\psi_-,-\psi_+),
\end{equation}
that is,
\begin{subequations}
\begin{align}
\delta \lambda &= \ell(\ell+1)a, \\
\delta a&= -\frac{\lambda}{\ell(\ell+1)}.
\end{align}
\end{subequations}
Since \cref{eq:L-EM-aux} is dynamically equivalent to the original Maxwell action \eqref{eq:L-EM}, related by auxiliary variables, a symmetry of one is a symmetry of the other. To construct the symmetry operators $\delta A_a$ and $\delta a$ for \cref{eq:L-EM} we need only use \cref{eq:Aasol,eq:lambdasol} relating $A_a$ and $\lambda$ on shell to find
\begin{subequations}\label{eq:delta-Aa}
\begin{align}
\delta A_a &= \epsilon_{ab}\partial^b a, \\
\delta a &=-\frac{r^2}{\ell(\ell+1)}\epsilon^{ab}\partial_aA_b.
\end{align}
\end{subequations}
This is an \emph{off-shell} symmetry of the action \eqref{eq:L-EM}.
As discussed in the introduction, this symmetry is non-local. This is reflected in the transformation law for $a$, which contains the inverse spherical Laplacian in the form $1/\ell(\ell+1)$.\footnote{Recalling that $-\ell(\ell+1)$ is the eigenvalue of the spherical Laplacian $D^2$ for scalar spherical harmonics, we see that $D^2(\delta a \, Y) = r^2\epsilon^\ab \partial_a A_b \,Y$.} Interestingly the symmetry transformation for $A_a$ is local.

The transformation law \eqref{eq:delta-Aa} has a natural interpretation in terms of Hodge duality. Consider the dual field strength tensor,
\begin{equation}
\star F_\mn = \frac12\epsilon_{\mn\alpha\beta}F^{\alpha\beta}.
\end{equation}
The Maxwell equation is $\dd\star F=0$, so that on-shell $\star F = \dd \tilde A$ can be expressed in terms of a dual potential $\tilde A_\mu$. It turns out that the off-shell duality transformation $\delta A_\mu$ is just such a dual potential, that is,
\begin{equation}
\delta A_\mu = \tilde A_\mu
\end{equation}
where
\begin{equation}
\tilde A_\mu \dd x^\mu = \epsilon_{ab}\partial^b a(x) Y(\theta)\dd x^a -\frac{r^2}{\ell(\ell+1)}\epsilon^{ab}\partial_aA_b(x) B_A(\theta)\dd \theta^A
\end{equation}
solves
\begin{equation}
\star F_\mn = \partial_\mu\tilde A_\nu-\partial_\nu\tilde A_\mu
\end{equation}
on shell. The fact that $A_\mu$ and $\tilde A_\mu$ are related by integration, $\star\dd A = \dd \tilde A$, underlies the non-local nature of $\delta A_\mu$.

If we further package the electric and magnetic master variables into a complex scalar,
\begin{equation}
\psi\equiv\frac{\psi_+ - i\psi_-}{\sqrt 2},
\end{equation}
then the action \eqref{eq:L-EM-aux} is simply
\begin{equation}
S = \int\dd^2x\sqrt{-g}\left[-\partial_a\psi\partial^a\bar\psi-\frac{\ell(\ell+1)}{r^2}\psi\bar\psi\right]. \label{eq:L-EM-psi}
\end{equation}
Electric-magnetic duality acts as $\delta\psi=i\psi$, which is manifestly a symmetry. It is straightforward to obtain the conserved current via the standard Noether procedure,
\begin{align}
J_a &= i(\bar\psi\partial_a\psi - \psi\partial_a\bar\psi) \nonumber\\
&= \psi_+\partial_a\psi_- - \psi_-\partial_a\psi_+.
\end{align}

Intriguingly, the complex master field $\psi$, which we obtained by integrating out non-dynamical fields and canonically normalizing, turns out to be proportional to $(\ell,m)$ modes of the middle Newman-Penrose scalar $\phi_1=(1/2)(F_{ln}-F_{m\barm})$,
\begin{equation}
\sqrt\frac{\ell(\ell+1)}{2}\psi_{\ell m} = r^2(\phi_1)_{\ell m}.
\end{equation}
For this reason, it will be illuminating to recontextualize the foregoing $2+2$ calculation in the GHP formalism.

\subsection{Maxwell in GHP}

Analogously to the Weyl tensor, the electromagnetic field strength tensor $F_\mn$ can be fully encoded in three complex Maxwell scalars,
\begin{equation}
\phi_0=F_{lm}, \quad\phi_1=\frac12\left(F_{ln}-F_{m\barm}\right),\quad\phi_2=F_{\barm n},
\end{equation}
of GHP types $\{2,0\}$, $\{0,0\}$, and $\{-2,0\}$, respectively. We remind the reader of the notation $F_{lm}=F_\mn l^\mu m^\nu$, etc. The Maxwell Lagrangian is
\begin{align}
\mL &= -\frac14F_\mn F^\mn \nonumber\\ &= -(-l^{(\mu}n^{\nu)}+m^{(\mu}\barm^{\nu)})(-l^\alpha n^\beta+m^\alpha \barm^\beta)F_{\mu \alpha}F_{\nu\beta} \nonumber\\
&= \phi_1^2-\phi_0\phi_2 + \mathrm{c.c.} \label{eq:NP-maxwell}
\end{align}
Now we introduce an auxiliary complex scalar $\lambda$ of GHP type $\{0,0\}$, meant to equal $\phi_1$ on-shell, by sending $\mL \to \mL - (\phi_1-\lambda)^2- (\bar\phi_1-\bar\lambda)^2$,
\begin{equation}
\boxed{\mathcal{L} = 2\phi_1\lambda-\lambda^2-\phi_0\phi_2+\mathrm{c.c.} \label{eq:LauxEM-GHP}}
\end{equation}

Instead of decomposing $A_\mu$ into $\mathcal{M}^2$ tensors $A_a(t,r)$ and $a(t,r)$ as in the $2+2$ decomposition, in the GHP formalism we encode it in the four scalars $(A_l,A_n,A_m,A_{\bar m})$.
The gauge choice we made earlier can be written in a GHP-invariant manner as
\begin{equation}
\eth'A_m+\eth A_{\bar m}=0.
\end{equation}
In this gauge, the even modes live in $A_l$ and $A_n$ while the odd modes live in $A_m$ and $A_\barm$ through the combination
\begin{equation}
\eth A_{\bar m} -\eth'A_m=i\frac{\ell(\ell+1)}{r^2}aY.
\end{equation}

To work with the equations of motion coming from the Lagrangian \eqref{eq:LauxEM-GHP}, it is helpful to establish just a bit more notation. First, we write the Maxwell scalars in terms of operators $\mathcal{T}_i$ acting on $A$ \cite{Wardell:2020naz},
\begin{subequations}
\begin{align}
\phi_0 = \mathcal{T}_0A &= -\eth A_l+(\tho-\rho)A_m,\\
\phi_1 = \mathcal{T}_1A &= \frac12\left(-\tho'A_l + \tho A_n + \eth' A_m - \eth A_\barm\right),\\
\phi_2 = \mathcal{T}_2A &= \eth' A_n-(\tho'-\rho')A_\barm.
\end{align}
\end{subequations}
Second, we introduce Wald's notion of \emph{adjoint operators} \cite{Wald:1978vm}. The adjoint $\mathcal{O}^\dag$ of an operator $\mathcal{O}$ satisfies $A\mathcal{O}B-B\mathcal{O}^\dag A=\nabla_\mu v^\mu$ for some vector $v^\mu$ and tensors (with indices suppressed) $A$ and $B$, so that under an integral we obtain the adjoint when integrating by parts,
\begin{equation}
\int\dd^4x\sdg A \mathcal{O} B = \int\dd^4x\sdg B \mathcal{O}^\dag A.
\end{equation}
The adjoints of the GHP derivatives are
\begin{equation}
\tho^\dag = -\tho+2\rho,\quad \eth^\dag = -\eth, \label{eq:GHPadjoints}
\end{equation}
along with their primes.
The adjoints of $\mathcal{T}^i$ are \cite{Wardell:2020naz}
\begin{subequations}
\begin{align}
\mathcal{T}_0^\dag &= l^\mu\eth - m^\mu(\tho-\rho),\\
\mathcal{T}_1^\dag &= \frac12\left[l^\mu(\tho'-2\rho')-n^\mu(\tho-2\rho)-m^\mu\eth'+\barm^\mu\eth\right],\\
\mathcal{T}_2^\dag &= -n^\mu\eth' +\barm^\mu(\tho'-\rho').
\end{align}
\end{subequations}

We now have the tools to vary the Maxwell Lagrangian \eqref{eq:LauxEM-GHP} with respect to $A$,
\begin{equation}
\left(\mathcal{T}_0^\dag \mathcal{T}_2+\mathcal{T}_2^\dag  \mathcal{T}_0\right)A+ \mathrm{c.c.} = 2 \mathcal{T}_1^\dag\lambda + \mathrm{c.c.}
\end{equation}
Note that this is a vector-valued equation, per the definitions of $\mathcal{T}_i^\dag$. The components along $l$ and $n$ determine $A_l$ and $A_n$ in terms of $\lambda$ and its complex conjugate,
\begin{subequations}
\begin{align}
A_l &= -\frac{1}{2\eth\eth'}(\tho-2\rho)(\lambda+\bar\lambda-\mathfrak{g}) \\
A_n &= \frac{1}{2\eth\eth'}(\tho'-2\rho')(\lambda+\bar\lambda+\mathfrak{g})
\end{align}
\end{subequations}
where
\begin{equation}
\mathfrak{g} \equiv \eth'A_m+\eth A_{\bar m}
\end{equation}
is zero in the gauge used in the previous subsection; we will fix $\mathfrak g=0$ herein. We can also integrate out $A_m$ and $A_\barm$ using the imaginary part of the $\lambda$ equation of motion,
\begin{align}
\lambda-\bar\lambda &= \phi_1-\bar\phi_1 \nonumber\\
&= \eth'A_m-\eth A_\barm,
\end{align}
which implies
\begin{equation}
A_m = \frac{1}{2\eth\eth'}\eth(\lambda-\bar\lambda),\qquad A_\barm = -\frac{1}{2\eth\eth'}\eth'(\lambda-\bar\lambda).
\end{equation}

Now that we have solutions for each component of $A_\mu$ in terms of $\lambda$, we can plug them into the Lagrangian \eqref{eq:LauxEM-GHP} to find a theory for $\lambda$ alone. However, to avoid the complications of dealing with the inverse $\eth\eth'$ operator, we first perform a simple field redefinition,
\begin{equation}
\lambda = \eth\eth' \psi
\end{equation}
so that the solution for $A_\mu$ is
\begin{subequations}
\begin{align}
A_l &= -\frac{1}{2}\tho(\psi+\bar\psi),\label{eq:Alsol}\\
A_n &= \frac{1}{2}\tho'(\psi+\bar\psi),\label{eq:Ansol} \\
A_m &= \frac{1}{2}\eth(\psi-\bar\psi)\label{eq:Amsol}.
\end{align}
\end{subequations}
To integrate out $A_\mu$ we plug this solution into \cref{eq:LauxEM-GHP}. The Maxwell scalars evaluated on this solution are
\begin{subequations}
\begin{align}
\phi_0 &= \eth\tho\psi,\\
\phi_1 &= \frac12\left[\tho\tho'\left(\psi+\bar\psi\right)+\eth\eth'(\psi-\bar\psi)\right],\\
\phi_2 &= \eth'\tho'\psi.
\end{align}
\end{subequations}
Putting these in the action we find, freely integrating by parts,\footnote{It is helpful to recall the GHP commutators \eqref{eq:GHPcommutators}, particularly $[\tho,\eth] = \rho\eth$. On GHP type $\{0,0\}$ objects such as $\psi$, $[\tho,\tho']=[\eth,\eth']=0$. Together with the adjoints \eqref{eq:GHPadjoints}, these imply that up to total derivatives $(\eth\tho A)(\eth'\tho'B)=(\tho\tho'A)(\eth\eth'B)=(\eth\eth'A)(\tho\tho'B)$.}
\begin{align}
\mathcal{L} &= 2\phi_1(\eth\eth'\psi)-(\eth\eth'\psi)^2-\phi_0\phi_2+\mathrm{c.c.} \nonumber\\
&= \left(\tho\tho'\psi-\eth\eth'\psi\right)\eth\eth'\bar\psi + \cc \nonumber\\
&= 2\left(\tho\tho'\psi-\eth\eth'\psi\right)\eth\eth'\bar\psi.
\end{align}
This is a remarkably simple result. To switch back to $\lambda = \eth\eth'\psi$, we integrate by parts and use the GHP commutators,
\begin{equation}
\eth\eth'(\tho\tho'-\eth\eth') = \left[(\tho-2\rho)(\tho'-2\rho') - \eth\eth'\right]\eth\eth',
\end{equation}
to write
\begin{equation}
\boxed{\mL = 2\bar\psi\left[(\tho'-2\rho')(\tho-2\rho)-\eth\eth'\right]\lambda.}\label{eq:Maxwell-GHP}
\end{equation}
where $\bar\psi = (\eth\eth')^{-1}\bar\lambda$. The equation of motion obtained by varying with respect to $\bar\psi$ is
\begin{equation}
\left[(\tho'-2\rho')(\tho-2\rho)-\eth\eth'\right]\lambda = 0.
\end{equation}
On shell $\lambda=\phi_1$, for which this is the \emph{Fackerell-Ipser equation} \cite{Fackerell:1972hg} in GHP notation \cite{Bini:2002jx}.

Electric-magnetic duality transformations act as complex rotations on the Maxwell scalars, $\phi_i \to e^{i\theta}\phi_i$, essentially since they are the components of the (anti-)self-dual parts of the Maxwell tensor. The action \eqref{eq:Maxwell-GHP} is indeed manifestly invariant under $\lambda\to e^{i\theta}\lambda$, or infinitesimally $\delta\lambda=i\lambda$ (along with $\delta\bar\psi=-i\bar\psi$).\footnote{The Lagrangian \eqref{eq:Maxwell-GHP} does not look real, but it is up to a total derivative, as can be explicitly checked using the commutators and adjoints of the GHP derivatives, and in particular the identity \[ \tho^\dag (\tho')^\dag \eth\eth' = \eth\eth'\tho\tho'. \]}

A natural extension of the setup with $\phi_1$ as an auxiliary field is to introduce auxiliary fields for all three Maxwell scalars, that is, a triplet $(\lambda_0,\lambda_1,\lambda_2)$ which on-shell satisfy $\lambda_i=\phi_i$.\footnote{This is essentially the construction of, e.g., \rcite{Krasnov:2022mvn} for the chiral formulation of Maxwell theory.} First let us note that we can ``chop off'' the $+\cc$ in the real Maxwell Lagrangian \eqref{eq:NP-maxwell} by adding the total derivative $(i/4)F_\mn(\star F)^\mn = \phi_0\phi_2-\phi_1^2-\cc$,
\begin{align}
\mL &= -\frac14F_\mn^2 + \frac i4 F_\mn (\star F)^\mn \nonumber\\
&= 2\left(\phi_1^2-\phi_0\phi_2\right).
\end{align}
Now we add in the full triplet of auxiliary fields,
\begin{align}
\mL &\to \mL - 2(\phi_1-\lambda_1)^2 + 2(\phi_0-\lambda_0)(\phi_2-\lambda_2) \nonumber\\
&= 2\left(2\phi_1\lambda_1 - \lambda_0\phi_2 - \lambda_2\phi_0 -\lambda_1^2 + \lambda_0\lambda_2\right).
\end{align}
The $\lambda_i$ equations of motion set $\lambda_i=\phi_i$ as desired, while the $A$ equation of motion is
\begin{equation}
2\mathcal{T}_1^\dag\lambda_1 - \mathcal{T}^\dag_0\lambda_2 - \mathcal{T}^\dag_2\lambda_0 = 0,
\end{equation}
or in vector notation,
\begin{align}
0 &= l^\mu \left[(\tho'-2\rho')\lambda_1-\eth\lambda_2\right] + m^\mu\left[(\tho-\rho)\lambda_2 - \eth'\lambda_1\right]  \nonumber\\
&- n^\mu \left[(\tho-2\rho)\lambda_1-\eth'\lambda_0\right]- \barm^\mu\left[(\tho'-\rho')\lambda_0-\eth\lambda_1\right] \nonumber\\
&\equiv \mathcal E_ll^\mu+\mathcal E_nn^\mu + \mathcal E_mm^\mu + \mathcal E_\barm \barm^\mu \label{eq:TSeom}.
\end{align}
This formulation yields first-order constraints among the $\phi_i$ on-shell. These are equivalent to the \emph{Teukolsky-Starobinsky identities}, which are second-order differential relations between $\phi_0$ and $\phi_2$, or equivalently fourth-order relations for $\phi_0$ and $\phi_2$ separately. To obtain the Teukolsky-Starobinsky identities we therefore need to take combinations of derivatives of $\mathcal{E}_a$ to remove $\phi_1$. The correct combinations are
\begin{subequations}
\begin{align}
\eth'\mathcal{E}_n - (\tho-3\rho)\mathcal{E}_m &=0 & \Longrightarrow && (\tho-2\rho)(\tho-2\rho)\phi_2 &= \eth'^2\phi_0,\\
\eth\mathcal{E}_l - (\tho-3\rho)\mathcal{E}_\barm &=0  & \Longrightarrow && (\tho'-2\rho')(\tho'-2\rho')\phi_0 &= \eth^2\phi_2,\\
\eth\mathcal E_m + \eth'\mathcal E_\barm &=0  & \Longrightarrow && (\tho'-2\rho')\eth'\phi_0 &= (\tho-2\rho)\eth\phi_2,
\end{align}
\end{subequations}
where the T-S identities following the arrows can be found in, e.g., eq. 43 of \rcite{Wardell:2020naz}. The third identity can also be obtained from $(\tho-2\rho)\mathcal E_l + (\tho'-2\rho')\mathcal E_n$. Here we have used the background equation $\tho\rho'=\tho'\rho=\rho\rho'-\Psi_2$.

We note that $\phi_1$ is special not just because it appeared naturally in the dynamical construction of the previous subsection, but also because it is closely related to the \emph{Killing-Yano 2-form} and its dual,
\begin{equation}
\phi_1 = \frac{i}{4r} F^\mn\left(Y_\mn - i \star Y_\mn\right),
\end{equation}
where
\begin{equation}
Y = r^3\sin\theta\dd\theta\wedge\dd\phi,\quad \star Y = r\dd t \wedge\dd r.
\end{equation}
The Killing tensor, which underlies separability, is the square of the Killing-Yano tensor, in coordinates,
\begin{equation}
k_\AB = -r^4\Omega_\AB,\quad k_{a \mu} = 0.
\end{equation}

To connect explicitly to the $2+2$ formulation of the previous subsection, we note the useful identities
\begin{subequations}
\begin{align}
m_A\bar m_B &= \frac{r^2}{2}\left(\Omega_\AB - i\epsilon_\AB\right), \\
l_an_b&=\frac12\left(-g_\ab+\epsilon_\ab\right).
\end{align}
\end{subequations}
Using these we can calculate the Maxwell scalars in terms of $2+2$ quantities,
\begin{subequations}
\begin{align}
\phi_1 &= \frac12\left(\epsilon^\ab\nabla_aA_b - i \frac{\ell(\ell+1)}{r^2}a\right)Y,\\
\phi_0 &= \left(\partial_aaB_A-A_aE_A\right)l^am^A, \\
\phi_2 &= -\left(\partial_aaB_A-A_aE_A\right)n^a\bar m^A,\\
\phi_0\phi_2 &= \frac14\left(A_a^2E_A^2+(\partial a)^2B_A^2\right).
\end{align}
\end{subequations}

We conclude with speculation about the structure discussed in this section and its generalization to Kerr. There the Fackerell-Ipser equation is not separable, which is why it is typical to work with the \emph{Teukolsky equations} \cite{Teukolsky:1972my} for the extreme-weight scalars $\phi_0$ and $\phi_2$, which are separable due to the aforementioned Killing tensor structure \cite{Frolov:2017kze}. It would be very interesting to obtain an action principle for the Teukolsky equations analogously to the one we have constructed for the Fackerell-Ipser equation and Teukolsky-Starobinsky identities.
We note that in \rcite{Toth:2018qrx} such an action was constructed using the fact that the Teukolsky equations are linear, which may provide a hint: the Teukolsky Lagrangian derived there is of the form $\mL\sim\rho^{-2}\phi_2\mathcal{O}\phi_0$, where $\mathcal{O}$ is the Teukolsky operator for $\phi_0$.
It would also be interesting to understand how the Debye and Hertz potentials which appear in reconstruction methods \cite{Cohen:1974cm,Wald:1978vm,Stewart:1978tm,Dolan:2019hcw} arise from the action formulation. We leave these important open questions for future work.

\section{Gravity}
\label{sec:gravity}

Consider linear perturbations around the Schwarzschild metric $\bar g_\mn$,\footnote{For black hole perturbation theory in $2+2$ language see, e.g., \rcite{Sarbach:2001qq,Martel:2005ir,Chaverra:2012bh,Ripley:2017kqg}. The factor of $2/\Mp$ is to canonically normalize the metric fluctuation.}
\begin{equation}
g_\mn = \bar g_\mn+\frac2\Mp h_\mn,
\end{equation}
and expand the Einstein-Hilbert action to quadratic order in $h_\mn$,
\begin{align}
S &= \frac{\Mp^2}{2}\int\dd^4x\sqrt{-g}R[g] \nonumber\\&= \bar S+\delta_1S+\delta_2S+\mathcal{O}(h^3). \label{eq:EH-NL}
\end{align}
The even- and odd-parity perturbations decouple at this order, so each is described by a separate quadratic action:
\begin{equation}
\delta_2S = \displaystyle\sum_{\ell=2}^\infty\displaystyle\sum_{m=-\ell}^\ell \left(S^{\ell m}_\mathrm{even}+S_\mathrm{odd}^{\ell m}\right).
\end{equation}
Herein we will drop bars on background quantities, since we will only be interested in $\delta_2S$.

Expanding the Ricci scalar to second order in perturbations is a non-trivial task, and ultimately not necessary, since we can write the action in first-order form. To see this, consider a metric variation $g\to g+\delta g$ and Taylor expand the action,
\begin{equation}
S[g+\delta g] = S[g] + \delta S + \frac12\delta^2S+\cdots.
\end{equation}
Matching to \cref{eq:EH-NL} we see that
\begin{equation}
\delta_2S = \frac12\delta^2S.
\end{equation}
It is a foundational result in GR that $\delta \int\dd^4x\sdg R = \int\dd^4x\sdg G_\mn \delta g^\mn$. Taking a second variation we obtain
\begin{align}
\delta_2 S &= \frac{\Mp^2}{4}\int\dd^4x\sdg \delta G_\mn \delta g^\mn \nonumber\\
&= -\int\dd^4x \sdg h^\mn G[h]_\mn \label{eq:quad-EH}
\end{align}
where
$G[h]_\mn\equiv\delta G_\mn[g+h]$ is the linear-in-$h$ part of the Einstein tensor for $g_\mn+h_\mn$,
\begin{equation}
G[h]_\mn = \nabla_\alpha \nabla_{(\mu}h_{\nu)}^\alpha - \frac12\Box h_\mn - \frac12\nabla_\mu\nabla_\nu h - \frac12\left(\nabla_\mu\nabla_\nu h^\mn - \Box h\right)g_\mn.
\end{equation}
For simplicity (and to facilitate comparison to the literature) we will continue to call this $\delta G_\mn$, with the understanding that it is evaluated on $g_\mn + h_\mn$ rather than $g_\mn + 2\Mp^{-1}h_\mn$. Integrating by parts we recover the standard Fierz-Pauli Lagrangian for a spin-2 field,
\begin{equation}
\delta_2S = \int\dd^4x \sdg\left(-\frac12\nabla_\alpha h_\mn\nabla^\alpha h^\mn + \nabla_\alpha h_\mn \nabla^\nu h^{\mu\alpha} - \nabla_\mu h \nabla_\nu h^\mn  + \frac12\nabla_\mu h\nabla^\mu h \right).
\end{equation}

The $2+2$ components of $\delta G_\mn[g+h]$ are standard and can be found in, e.g., \rcite{Sarbach:2001qq,Martel:2005ir,Chaverra:2012bh}.\footnote{We leave the analogous GHP analysis for future work.} We present relevant components in \cref{app:22Ricci}. The quadratic action \eqref{eq:quad-EH} is expanded as
\begin{equation}
h^\mn \delta G_\mn = h^\ab \delta G_\ab + \frac{2}{r^2}h^{aA}\delta G_{aA} + \frac{1}{r^4}h^\AB\delta G_\AB.
\end{equation}
We remind the reader that $\mathcal{M}_2$ indices are raised with $g^\ab$ and $S^2$ indices with $\Omega^\AB$.

There are at least two useful gauges which can be safely fixed at the level of the action \cite{Motohashi:2016prk}. One is the standard Regge-Wheeler gauge, in which $h_{aA}$ is purely odd and $h_\AB=r^2K\Omega_\AB$. Another is the ``$\alpha$ gauge'' used in, e.g., \rcite{DeFelice:2011ka,Kobayashi:2014wsa,Hui:2020xxx}, where $h_{aA}$ contains both even and odd pieces and $h_\AB=0$. The gauge choice affects the auxiliary structure of the action. To see this, consider the gauge-invariant variables $\tilde h_\ab$ and $\tilde K$ defined in \rcite{Martel:2005ir}, which correspond (by construction) to $h_\ab$ and $K$ in the Regge-Wheeler gauge, and in $\alpha$ gauge contain derivatives,
\begin{subequations}
\begin{align}
\tilde h_\ab &= h_\ab - 2\nabla_{(a}\left(r^2r_{b)}\alpha\right),\\
\tilde K &= -2fr\alpha.
\end{align}
\end{subequations}
We will remain agnostic about which of these two gauges to pick, and write down expressions for both.

In these gauges, the components of $h_\mn$ are
\begin{subequations}
\label{eq:gauge-h}
\begin{align}
h_\ab &= \displaystyle\sum_{\ell,m}h^{\ell m}_\ab Y_{\ell m},\\
h_{aA} &= \displaystyle\sum_{\ell,m}r^2\left(\alpha_{\ell m}r_aE_A^{\ell m}  + h^{\ell m}_a B_A^{\ell m}\right),\label{eq:haA}\\
h_\AB &= \displaystyle\sum_{\ell, m}r^2K_{\ell m}Y_{\ell m}\Omega_\AB,
\end{align}
\end{subequations}
where we remind the reader that $r_a\equiv\partial_a r$. As usual we will drop the summation and the subscripts and focus on a single $(\ell,m)$ mode. In Regge-Wheeler gauge we set $\alpha=0$, and in $\alpha$ gauge we set $K=0$.
We will also find it convenient to decompose $h_\ab$ into its trace and tracefree parts,
\begin{equation}
h_\ab = \hat h_\ab + \frac12hg_\ab,\quad \hat h^a{}_a = 0,
\end{equation}
and to work with the Ricci tensor rather than the Einstein tensor,
\begin{equation}
\delta G_\mn = \delta R_\mn - \frac12\delta R g_\mn,\quad\delta R = g^\ab\delta R_\ab + r^{-2}\Omega^\AB \delta R_\AB
\end{equation}
In terms of these variables, the even and odd actions are
\begin{subequations}
\begin{align}
S_\mathrm{even} &= \int\dd^2x \sdg \dd\Omega \left[r^2 (Kg^\ab-\hat h^\ab) \delta R_\ab + \frac12 h \Omega^\AB\delta R_\AB  -2 r^2 r^a \alpha E^A\delta R_{aA}\right], \\
 S_\mathrm{odd} &=  -2\int\dd^2x \sdg \dd\Omega r^2 h^a B^A\delta R_{aA}.
\end{align}
\end{subequations}

To integrate over the 2-sphere, we note that the $S^2$ scalars $\delta R_\ab$ and $\Omega^\AB \delta R_\AB$ are expanded in $Y_{\ell m}$, while the even and odd parts of $\delta R_{aA}$ can be written as
\begin{equation}
\delta R_{aA} = \delta R^E_a E_A + \delta R^B_a B_A.
\end{equation}
Performing the integral over $S^2$ and writing the actions as $S=\int\dd^2x\sdg\mL$, the Lagrangians are
\begin{subequations}
\begin{align}
\mL_\mathrm{even} &= r^2 (Kg^\ab-\hat h^\ab) \delta R_\ab + \frac12 h \Omega^\AB\delta R_\AB  -2\ell(\ell+1)r^2 r^a\alpha  \delta R^E_{a},\label{eq:action-even} \\
\mL_\mathrm{odd} &=  -2\ell(\ell+1)  r^2 h^a\delta R^B_{a}, \label{eq:action-odd}
\end{align}
\end{subequations}
where $h_\ab$ denotes $h_\ab^{\ell m}$, etc. Let us treat the odd and even sectors separately.

\subsection{Odd sector}

The odd piece of the Ricci tensor is (see \cref{app:22Ricci})
\begin{equation}
\delta R^B_a = \frac{1}{2r^2}\nabla^b\left(r^4F_\ab\right) + \frac{(\ell+2)(\ell-1)}{2}h_a,
\end{equation}
where
\begin{equation}
F_\ab = \partial_ah_b-\partial_bh_a,
\end{equation}
so the Lagrangian \eqref{eq:action-odd} is
\begin{align}
\mL_\mathrm{odd} &=  -2\ell(\ell+1)  r^2 h^a\delta R^B_{a} \nonumber\\
&= -\ell(\ell+1)\left(h^a\nabla^b\left(r^4F_\ab\right) + (\ell+2)(\ell-1)r^2h_a^2\right) \nonumber\\
&= -\ell(\ell+1)\left(\frac12r^4F_\ab^2 + (\ell+2)(\ell-1)r^2h_a^2\right),
\end{align}
where in the last line we have integrated by parts. Note that $(\ell+2)(\ell-1)=\ell(\ell+1)-2$. Finally we rescale
\begin{equation}
h_a \to \frac{h_a}{\sqrt{2\ell(\ell+1)}}
\end{equation}
so the action takes the form
\begin{equation}
\boxed{\mL_\mathrm{odd} = -\frac14r^4F_\ab^2 - \frac{(\ell+2)(\ell-1)}{2}r^2h_a^2.} \label{eq:odd-action-22}
\end{equation}
In coordinates this is \cite{Hui:2020xxx}
\begin{equation}
\mathcal{L}_\mathrm{odd} = \frac12r^4(\dot{h}_1 - h_0')^2 +\frac12(\ell+2)(\ell-1)r^2\left(\frac1fh_0^2 - f h_1^2\right),\label{eq:odd-action-22-coords}
\end{equation}
where $h_a\dd x^a = h_0\dd t + h_1\dd r$, and overdots and primes denote $\partial_t$ and $\partial_r$, respectively.

Physically we can think of \cref{eq:odd-action-22} as describing a two-dimensional vector with an $r$-dependent mass,\footnote{The physical intuition behind this is a bit more apparent if we perform the $2+2$ decomposition before linearizing. In Regge-Wheeler gauge the odd modes only contribute to the Ricci scalar via the tangent and transverse extrinsic curvatures, in the form of a mass term and a Maxwell term, respectively, for $g^{AB} B_B\delta g_{aA}$ \cite{Ripley:2017kqg}. The factor of $r^{-2}$ in $g^\AB=r^{-2}\Omega^\AB$ motivates the compensating factor of $r^2$ in our definition \eqref{eq:haA} of $\delta g_{aA}$.} where we remind the reader that $r$ is a background scalar rather than necessarily a coordinate direction.

Note the close resemblance to $\mL_\mathrm{even}$ for the Maxwell field \eqref{eq:L-EM}. We can repeat the same trick to integrate out the two fields $h_a$ in favor of a single dynamical field. We integrate in an auxiliary variable $\lambda(x^a)$ via a perfect square so as not to affect the dynamics,
\begin{align}
\mathcal{L}_\mathrm{odd,aux} &= \mathcal{L}_\mathrm{odd} +\frac14\left(r^2 F_{ab}+\lambda \epsilon_{ab}\right)^2\nonumber\\
&= \frac12\left(r^2\lambda\epsilon^\ab F_\ab - \lambda^2 - (\ell+2)(\ell-1) r^2h^2\right).
\end{align}
This is dynamically equivalent to $\mL_\mathrm{odd}$, which is recovered by plugging in the solution to the $\lambda$ equation of motion, $\lambda = (1/2)r^2\epsilon^\ab F_\ab$, and we will write it as $\mL_\mathrm{odd}$ accordingly. The introduction of $\lambda$ gives us the option to integrate out $h_a$ by solving its equation of motion,
\begin{equation}
(\ell+2)(\ell+1)r^2h_a = \epsilon_\ab \partial^b (r^2\lambda). \label{eq:ha-sol}
\end{equation}
Substituting this into the action we have
\begin{equation}
\mathcal{L}_\mathrm{odd} = -\frac12\frac{[\partial(r^2\lambda)]^2}{(\ell+2)(\ell-1)r^2}-\frac12\lambda^2.
\end{equation}
We perform a further rescaling to canonically normalize the kinetic term,
\begin{equation}
\lambda = \frac{\sqrt{(\ell+2)(\ell-1)}}{r}\Psi_-,
\end{equation}
so that the action becomes, using the background equations of motion \eqref{eq:back-eom},
\begin{equation}
\mathcal{L}_\mathrm{odd} = -\frac12(\partial\Psi_-)^2 - \frac12\left(\frac{\ell(\ell+1)}{r^2}-\frac32 R\right)\Psi_-^2.
\end{equation}
The mass term explicitly evaluates to
\begin{align}
\frac{\ell(\ell+1)}{r^2}-\frac32R&= \frac{\ell(\ell+1)}{r^2}-\frac{3\rs}{r}\nonumber\\& = \frac{V_-(r)}{f(r)},\label{eq:VRW}
\end{align}
where $V_-(r)$ is the \emph{Regge-Wheeler potential} \cite{Regge:1957td}. Putting everything together we obtain the odd-sector Regge-Wheeler action,
\begin{equation}
\boxed{S_\mathrm{odd} = \int\dd^2x\sqrt{- g}\left[-\frac12(\partial\Psi_-)^2-\frac12\frac{V_-}{f}\Psi_-^2\right].}
\end{equation}

The equation of motion, 
\begin{equation}
\Box\Psi_- = \frac{V_-}{f}\Psi_-,
\end{equation}
where
\begin{align}
\Box &= \partial_a(g^\ab\partial_b) \nonumber\\
&= \frac{1}{f}\left(-\partial_t^2+\partial_{r_\star}^2\right),
\end{align}
is the usual Regge-Wheeler equation \cite{Regge:1957td} for $\Psi_-$,
\begin{equation}
\left(-\partial_t^2+\partial_{r_\star}^2\right)\Psi_- = V_-\Psi_-.
\end{equation}
This means that $\Psi_-$ must be proportional to the Regge-Wheeler variable up to time derivatives. Indeed, recalling Martel and Poisson's \cite{Martel:2005ir} gauge-invariant definition of the Cunningham-Price-Moncrief variable \cite{Cunningham:1978zfa}, which is itself a time integral of the original Regge-Wheeler variable \cite{Regge:1957td}, we find agreement with $\Psi_-$ up to a numerical factor:
\begin{align}
\Psi_\mathrm{CPM} &= \frac{r^3}{(\ell+2)(\ell-1)}\epsilon^\ab F_\ab \nonumber\\
&= \frac{2r}{(\ell+2)(\ell-1)}\lambda \nonumber\\
&= \frac{2}{\sqrt{(\ell+2)(\ell-1)}}\Psi_-.
\end{align}

We conclude the discussion of the odd sector by noting an interesting alternative approach discussed in, e.g., \rcite{Sarbach:2001qq}. Consider the $\mathcal{M}^2$ 1-form $h=h_a\dd x^a$. The action is
\begin{equation}
S_\mathrm{odd} = -\frac12\int\dd^2x\left(r^4F\wedge\star F + (\ell+2)(\ell-1)r^2h\wedge\star h\right),
\end{equation}
and the equation of motion is
\begin{equation}
\mathcal{E} = \mathcal{E}_a\dd x^a = -\star\dd(r^4\star F) - (\ell+2)(\ell-1)r^2 h.
\end{equation}
Taking a divergence by applying $\dd \star$, we find that the 1-form $r^2\star h$ is closed,
\begin{equation}
\dd(r^2\star h)=0.
\end{equation}
By the Poincar\'e lemma we can write it in terms of a scalar potential $\phi$,
\begin{equation}
r^2 h = \star\dd\phi, \label{eq:h-phi}
\end{equation}
or in index notation,
\begin{equation}
r^2h_a = -\epsilon_\ab \partial^b\phi.
\end{equation}
Comparing to \cref{eq:ha-sol} we see that this potential is related to our auxiliary variable $\lambda$ by
\begin{equation}
\phi = -\frac{r^2}{(\ell+2)(\ell-1)}\lambda.
\end{equation}

The auxiliary field method is a technique for consistently implementing \cref{eq:h-phi} at the level of the action. In particular, if we were to na\"ively plug the solution \eqref{eq:h-phi} directly into the original action \eqref{eq:odd-action-22}, the resulting theory would be of fourth order in derivatives of $\phi$, and could not describe the same physics: it contains two degrees of freedom rather than one, and possesses an Ostrogradski ghost instability \cite{Woodard:2015zca}.

\subsection{Even sector}

The action for the even sector is given by \cref{eq:action-even}. Expressions for relevant components of the perturbed Ricci tensor are in \cref{app:22Ricci}. The resulting actions after many intergrations by parts are
\begin{subequations}\label{eq:even-action-22}
\begin{align}
\mathcal{L}_\mathrm{even}^\mathrm{RW} &= -2rr^c\hat h^\ab \nabla_a\hat h_{bc} -2r\hat h^\ab r_a\partial_bh  - \frac{r^2R+\ell(\ell+1)+1}{2}\hat h_\ab^2      + \frac{\ell(\ell+1)+2}{4}h^2  \nonumber\\
&\hphantom{{}=}-2r^2\nabla_a\hat h^\ab\nabla_bK  +r^2\partial h\cdot\partial K + 2rr^aK\partial_ah+\ell(\ell+1)hK +  r^2(\partial K)^2,\label{eq:even-action-22-RW} \\
\mL_\mathrm{even}^\alpha &= - 2 r r^c\hat h^\ab\nabla_a\hat h_{bc}  -2r\hat h^\ab r_a\partial_bh   -\frac{r^2R+\ell(\ell+1)+1}{2}\hat h_\ab^2  + \frac{\ell(\ell+1)+2}{4}h^2 \nonumber\\
&\hphantom{{}=}+ \ell(\ell+1)r^2\left(r^2 (\epsilon^\ab r_a\partial_b\alpha)^2 + 2 f\alpha^2   -2r_b\alpha\nabla_a\hat h^\ab   -\frac{1}{r^4}h\nabla_a\left(r^4r^a\alpha\right)\right),\label{eq:even-action-22-alpha}
\end{align}
\end{subequations}
in Regge-Wheeler gauge and in $\alpha$ gauge, respectively.
We begin by noting the well-known fact that these expressions are significantly more complicated than \cref{eq:odd-action-22}.\footnote{One is compelled to wonder who ordered this, especially given that, as we will see, the two sectors have essentially the same dynamics.}

It is convenient to perform a coordinate-like decomposition on objects with indices by projecting along $r_a$ and the timelike direction $t_a=\epsilon_\ab r^b=f\partial_at$, in terms of which the metric is \cite{Martel:2005ir}
\begin{equation}
fg_\ab = r_ar_b-t_at_b.
\end{equation}
In particular, we do not lose any information by projecting the traceless perturbation $\hat h$ once along $r_a$ \cite{Chaverra:2012bh},
\begin{equation}
\hat h_a \equiv \hat h_\ab r^b,
\end{equation}
as we can reconstruct $\hat h_\ab$ via\footnote{To see this, consider all contractions with $r_a$ and $t_a$.}
\begin{equation}
f\hat h_\ab = 2r_{\langle a}\hat h_{b\rangle} = r_a\hat h_b+r_b\hat h_a - (r\cdot \hat h)g_\ab,
\end{equation}
where angular brackets denote traceless symmetrization, $T_{\langle\ab\rangle} = T_{(\ab)} - \frac12Tg_\ab$. This simplifies the actions somewhat,
\begin{subequations}\label{eq:h-hat-actions}
\begin{align}
\mathcal{L}_\mathrm{even}^\mathrm{RW} &= - 2r\hat h^\ab \nabla_a \hat h_b -\frac{\ell(\ell+1)+1}{f}\hat h_a^2   -2r\hat h^a\partial_ah   + \frac{\ell(\ell+1)+2}{4}h^2  \nonumber\\
&\hphantom{{}=}-2r^2\nabla_a\hat h^\ab\nabla_bK  +r^2\partial h\cdot\partial K + 2rr^aK\partial_ah+\ell(\ell+1)hK +  r^2(\partial K)^2, \\
\mL_\mathrm{even}^\alpha &= - 2r\hat h^\ab \nabla_a \hat h_b -\frac{\ell(\ell+1)+1}{f}\hat h_a^2 - 2r \hat h^a\partial_ah + \frac{\ell(\ell+1)+2}{4}h^2 \nonumber\\
&\hphantom{{}=}+ \ell(\ell+1)r^2\left( r^2 (t^a\partial_a\alpha)^2 + 2 f\alpha^2   -2r_b\alpha\nabla_a\hat h^\ab   -\frac{1}{r^4}h\nabla_a\left(r^4r^a\alpha\right)\right).
\end{align}
\end{subequations}

For concreteness, let us fix $\alpha$ gauge. We will discuss Regge-Wheeler gauge at the end of the section. After the gauge freedom has been used up, there are four fields for one underlying dynamical degree of freedom. Two auxiliary variables are apparent by inspection of the action \eqref{eq:h-hat-actions}: $t^a\hat h_a\sim h_{tr}$ and $h$. Here we will essentially follow the procedure of \rcite{Hui:2020xxx} and begin by integrating out the former.
To isolate the components of $\hat h_a$ we decompose it as
\begin{equation}
\hat h_a = \hat h_0 t_a + \hat h_1r_a. \label{eq:hhat-decomp}
\end{equation}
We will also need to perform some simple field redefinitions to demix fields. We begin by shifting $h$,
\begin{equation}
h = \tilde h - 2\hat h_1.
\end{equation}
Note that $h$ contains both $h_{tt}$ and $h_{rr}$, whereas $\tilde h \sim h_{rr}$. In this field basis the action is
\begin{align}
\mL_\mathrm{even}^\alpha &= \ell(\ell+1)\hat h_0^2   + \frac{\ell(\ell+1)+2}{4}\tilde h^2 - \ell(\ell+1)\tilde h \hat h_1  -2rt^a\hat h_0 \partial_a\tilde h +2rr^a\tilde h\partial_a \hat h_1 \nonumber\\
&\hphantom{{}=}  -\frac{\ell(\ell+1)}{r^2}\tilde h\nabla_a\left(r^4r^a\alpha\right)   + \ell(\ell+1)r^2\left[ r^2 (t^a\partial_a\alpha)^2 + 2 f\alpha^2\right]  \nonumber\\
&\hphantom{{}=} + 2\ell(\ell+1)\left[\hat h_0 t^a \partial_a(r^2\alpha) + 2rr^a\hat h_1\partial_a(r\alpha) +  (1+3f)r\hat h_1\alpha\right]
\end{align}
We can integrate out $\hat h_0$ using its equation of motion,
\begin{equation}
\ell(\ell+1)\hat h_0 = rt^a \partial_a\left(\tilde h  -\ell(\ell+1)r\alpha\right),
\end{equation}
to find
\begin{align}
\mL_\mathrm{even}^\alpha &= -\frac{r^2}{\ell(\ell+1)}t^at^b \partial_a\tilde h\partial_b(\tilde h  -2\ell(\ell+1)r\alpha) + \frac{\ell(\ell+1)+2}{4}\tilde h^2 - \ell(\ell+1)\tilde h \hat h_1   \nonumber\\
&\hphantom{{}=} +2rr^a\tilde h\partial_a \hat h_1   -\frac{\ell(\ell+1)}{r^2}\tilde h\nabla_a\left(r^4r^a\alpha\right)    \nonumber\\
&\hphantom{{}=} + 2\ell(\ell+1)\hat h_1\left(2rr^a\partial_a(r\alpha) +  (1+3f)r\alpha\right) + 2\ell(\ell+1)r^2f\alpha^2.
\end{align}
Now we perform a second field redefinition,\footnote{The reason for this particular order of operations is that integrating out $\hat h_0$ simplifies the kinetic term for $\alpha$.} comprising a shift to demix $\alpha$ and $\tilde h$ and an overall rescaling,
\begin{equation}
\alpha = \frac{\Lambda}{r^2f}\psi + \frac{\tilde h}{2\ell(\ell+1)r},
\end{equation}
where we have introduced the function \cite{Martel:2005ir}
\begin{equation}
\Lambda(r) \equiv \ell(\ell+1) +1 -3f.
\end{equation}
The action becomes
\begin{align}
\mL_\mathrm{even}^\alpha &= \left(\frac{\ell(\ell+1) +1}{4} + \left(\frac{1}{2\ell(\ell+1) }-1\right)f\right)\tilde h^2 - \Lambda\tilde h \hat h_1   -\frac{\ell(\ell+1) \Lambda}{f}r^a\tilde h\partial_a\psi \nonumber\\
&\hphantom{{}=}  + \frac{2\Lambda r}{f}t^at^b \partial_a\tilde h\partial_b\psi  - \frac{(\ell+2)(\ell-1)(\ell(\ell+1)+\Lambda)}{r}\tilde h\psi + \frac{2\ell(\ell+1) \Lambda^2}{r^2f}\psi^2 \nonumber\\
&\hphantom{{}=} + \frac{2\ell(\ell+1)}{f}\hat h_1\left[2\Lambda r^a\partial_a\psi + \frac{3f(\ell(\ell+1)-f)-\ell(\ell+1)-1}{r}\psi\right] .
\end{align}
Note that $\psi$ is precisely the gauge-invariant Zerilli-Moncrief function defined in \rcite{Martel:2005ir}, multiplied by $-1/4$.

The upshot of all these field redefinitions is that two of the remaining three fields are manifestly non-dynamical: $\hat h_1$ is a Lagrange multiplier (it appears linearly) and $\tilde h$ is auxiliary (it appears quadratically but without derivatives). The constraint obtained by varying with respect to $\hat h_1$ fixes $\tilde h$ in terms of $\psi$,
\begin{equation}
\tilde h = \frac{2\ell(\ell+1)}{f}\left[2r^a\partial_a + \frac{3f(\ell(\ell+1)-f)-\ell(\ell+1)-1}{r\Lambda}\right]\psi, \label{eq:htildesol}
\end{equation}
while the equation of motion for $\tilde h$ is
\begin{align}
0 &= \Lambda\hat h_1 + \frac{(\ell(\ell+1)+1-f) \Lambda}{f}r^a\partial_a\psi + \frac{(\ell+2)(\ell-1)(\ell(\ell+1)+\Lambda)}{r}\psi \nonumber\\
&\hphantom{{}=} + \frac{2r\Lambda}{f}t^at^b \nabla_a\nabla_b\psi - \left(\frac{\ell(\ell+1) +1}{2} + \left(\frac{1}{\ell(\ell+1)}-2\right)f\right)\tilde h.
\end{align}
This fixes $\hat h_1$ once we use \cref{eq:htildesol}, although we do not need to know $\hat h_1$ in order to integrate it out of the action, as it multiplies the constraint \eqref{eq:htildesol} that it enforces. We will however need this equation in order to construct off-shell duality operators for the metric perturbations.

Plugging \cref{eq:htildesol} into the action we finally obtain, after some integrations by parts and algebra,
\begin{equation}
\mL_\mathrm{even}^\alpha = -4\ell(\ell+1)(\ell+2)(\ell-1)\left[(\partial\psi)^2 + \frac{V_+}{f}\psi^2\right],
\end{equation}
where
\begin{equation}
V_+ = \frac{f}{3r^2}\left(\Lambda + \frac{2(\ell+2)^2(\ell-1)^2\left(1+\ell(\ell+1)\right)}{\Lambda^2}\right)
\label{eq:VZ}
\end{equation}
is the \emph{Zerilli potential} \cite{Zerilli:1970se}.
Finally we canonically normalize,
\begin{equation}
\psi =  \frac{1}{2\sqrt{\ell(\ell+1)(\ell+2)(\ell-1)}}\Psi_+,
\end{equation}
to obtain the Zerilli action for the even sector:
\begin{equation}
\boxed{\mL_\mathrm{even} = -\frac12(\partial\Psi_+)^2 -\frac12 \frac{V_+}{f}\Psi^2_+.}\label{eq:actionZ}
\end{equation}

The main benefit of working with $\alpha$ gauge is that the field redefinitions we needed to perform did not involve derivatives, but a choice of gauge is not a choice of physics, and indeed in Regge-Wheeler gauge we can follow a similar procedure to reduce the action \eqref{eq:even-action-22-RW} to the Zerilli action \eqref{eq:actionZ}. We begin again by integrating out $h_{tr}\sim\hat h_0$, while $h_{tt}\sim h-2\hat h_1$ is a Lagrange multiplier that imposes a constraint on $K$ and $h_{rr}\sim h+2\hat h_1$ (and in turn drops out of the action). To demix the remaining two variables and canonically normalize we perform a field redefinition,
\begin{equation}
h+2\hat h_1 = \sqrt{\frac{\ell(\ell+1)}{2(\ell+2)(\ell-1)}}\frac{\Lambda}{rf}\Psi_+ + \frac1f\left(2rr^a\partial_a -\Lambda\right)K.
\end{equation}

The even sector is inordinately complicated, and the procedure we have done is not unique, and may not be the simplest or clearest. Alternative approaches would therefore be interesting to explore. An obvious alternative is to integrate out $h$ first rather than $\hat h_0$. Furthermore, the decomposition \eqref{eq:hhat-decomp} can be swapped for a more elegant argument in terms of differential forms analogously to the odd sector \cite{Chaverra:2012bh}, which may therefore admit an auxiliary variable formulation. And of course an approach eliding the Regge-Wheeler and Zerilli equations altogether in favor of the Teukolsky equation would be of exceptional interest.

\section{Chandrasekhar duality}
\label{sec:Chandra}

The linearized Einstein-Hilbert action is a complicated functional of the metric perturbations (cf. \cref{eq:odd-action-22,eq:even-action-22}), but by integrating out the non-dynamical degrees of freedom we obtained a simple action in terms of the Regge-Wheeler and Zerilli variables,
\begin{equation}\label{eq:EH-Sch}
\boxed{S =\displaystyle\sum_{\ell=2}^\infty\displaystyle\sum_{m=-\ell}^\ell\displaystyle\sum_\pm\int\dd^2x\sdg\left(  -\frac12(\partial\Psi_\pm)^2 -\frac{1}{2f}V_\pm\Psi^2_\pm\right),}
\end{equation}
where $V_+$ and $V_-$ are the usual Zerilli \cite{Zerilli:1970se} and Regge-Wheeler \cite{Regge:1957td} potentials, respectively.

It is important to pause here to emphasize the difference between on-shell and off-symmetries. We could have constructed \cref{eq:EH-Sch} directly from the Regge-Wheeler and Zerilli equations, but it was a non-trivial exercise to get there from the Einstein-Hilbert action using standard field theory tools. Having done this exercise, we will be able to construct an off-shell duality symmetry of the original action \eqref{eq:EH-NL}.

First let us demonstrate the duality invariance of the Regge-Wheeler/Zerilli action \eqref{eq:EH-Sch}. It is a remarkable fact that the two seemingly-disparate potentials $V_\pm$ (cf. \cref{eq:VRW,eq:VZ}) can be written in a unified form in terms of a single \emph{superpotential} \cite{Chandrasekhar:1975zza,1975RSPSA.343..289C,Chandrasekhar:1985kt,1984RSPSA.392....1C,Wald:1973wwa,Berti:2009kk},\footnote{Note also the relation $V_+ = V_- - 2\partial_{r_\star}^2\Lambda$ \cite{Aksteiner:2014zyp}. For electric-magnetic duality on charged black hole backgrounds, see \rcite{Pereniguez:2023wxf}.}
\begin{equation}
V_\pm = W^2 \mp r^a\partial_aW+\beta,
\end{equation}
where the superpotential $W(r)$ and constant $\beta$ are given by
\begin{equation}
W(r)= -\left(\frac32\frac{rRf}{\Lambda}+ \sqrt{-\beta}\right),\quad \beta \equiv -\left(\frac{\ell(\ell+1)(\ell+2)(\ell-1)}{6\rs}\right)^2.\label{eq:superpot}
\end{equation}
It is straightforward to check that the action \eqref{eq:EH-Sch} is invariant under the duality symmetry
\begin{equation}
\boxed{\delta\Psi_\pm = \left(r^a\partial_a\mp W\right)\Psi_\mp. }\label{eq:psisym}
\end{equation}
The transformation \eqref{eq:psisym} is an off-shell symmetry of the action, and coincides on shell with the venerable \emph{Chandrasekhar duality} \cite{Chandrasekhar:1975zza,1975RSPSA.343..289C,Chandrasekhar:1985kt,1984RSPSA.392....1C}.\footnote{For Chandrasekhar duality beautifully visualized, see \rcite{Nichols:2012jn}.} This ``hidden'' symmetry of the linearized Einstein equations relates a solution $\Psi_\pm$ to the Regge-Wheeler or Zerilli equation to a solution $\Psi_\mp$ to the other equation, which can be constructed in frequency space via\footnote{The prefactor is not strictly necessary since any constant multiple of this will also be a solution. However we include this prefactor to emphasize the existence of \emph{algebraically-special} modes for which $\omega^2=\beta$, where special care must be taken. We will not discuss algebraically-special modes in this work.}
\begin{equation}
\Psi_\pm = \frac{1}{\beta-\omega^2}\left(\frac{\dd}{\dd\rst}\Psi_\mp \mp W\Psi_\mp\right).\label{eq:Chandra}
\end{equation}
We note that, intriguingly, this symmetry structure also appears in \emph{supersymmetric quantum mechanics}, the theory of $0+1$-dimensional supersymmetry \cite{Cooper:1994eh}.\footnote{Concretely, we can write \cref{eq:psisym} in a manner suggestive of supersymmetric quantum mechanics by defining raising and lowering operators,
\begin{equation}
A \equiv \frac{\partial}{\partial r_\star} + W, \qquad A^\dag = -\frac{\partial}{\partial r_\star} + W,
\end{equation}
and writing the symmetry as
\begin{equation}
\delta\begin{pmatrix} 
\Psi_+ \\
\Psi_-
\end{pmatrix} = \begin{pmatrix} 
0 & -A^\dag \\
A & 0
\end{pmatrix} \begin{pmatrix} 
\Psi_+ \\
\Psi_-
\end{pmatrix}.
\end{equation}} The Chandrasekhar duality is responsible for the crucial result that, for four-dimensional black holes in GR, the even and odd sectors are \emph{isospectral}, meaning they share the same quasinormal mode spectrum.\footnote{This is a consequence of the fact that, if $\Psi_\pm$ satisfies the boundary conditions which define a quasinormal mode, then the $\Psi_\mp$ generated by \cref{eq:Chandra} does as well, so the Chandrasekhar transformation relates quasinormal modes of even and odd parity without changing the frequency (excluding algebraically-special modes). As we will see, this is also true for the infalling boundary conditions used to calculate Love numbers.}

With the off-shell symmetry \eqref{eq:psisym} in hand, we can compute conserved quantities using the Noether procedure. The conservation law, in coordinates, is
\begin{equation}
\partial_tJ^t + \partial_{r_\star}\jrs = 0,\label{eq:symcons}
\end{equation}
with the current
\begin{subequations}
\begin{align}
J^t &= \dot\Psi_+ A^\dag\Psi_- - \dot\Psi_-A\Psi_+ \nonumber \\
&= -\Psi_+'\dot\Psi_- - \dot\Psi_+\Psi_-' + W\left(\Psi_-\dot\Psi_+-\Psi_+\dot\Psi_-\right), \label{eq:Jt}\\
J^{r_\star} &= \dot\Psi_+\dot\Psi_--(A\Psi_+)(A^\dag\Psi_-)-\beta\Psi_+\Psi_- \nonumber\\
&= \dot\Psi_+\dot\Psi_- + \Psi_+'\Psi_-' + W\left(\Psi_+\Psi_-'-\Psi_-\Psi_+'\right) -\left(W^2+\beta\right)\Psi_+\Psi_-. \label{eq:Jrs}
\end{align}
\end{subequations}
Here overdots denote derivatives with respect to $t$, and primes denote $\partial_{r_\star}$ derivatives.

\subsection{A complex master variable}

Similarly to the spin-1 case, we can combine the Regge-Wheeler and Zerilli variables into a complex variable,
\begin{equation}
\Psi \equiv \frac{\Psi_+ + i\Psi_-}{\sqrt 2},
\end{equation}
in terms of which the Lagrangian \eqref{eq:EH-Sch} takes a very simple form,
\begin{align}
\mathcal{L} &= -\frac12\displaystyle\sum_\pm \left((\partial\Psi_\pm)^2+\frac{V_\pm}{f}\Psi_\pm^2\right) \nonumber\\
&= - \partial_a\Psi\partial^a\bar\Psi + \frac{1}{2f}r^a\partial_aW(\Psi^2+\bar\Psi^2) -\frac{W^2+\beta}{f}\Psi\bar\Psi
\end{align}
as does the duality transformation,
\begin{equation}
\delta\Psi = i\left(r^a\partial_a\bar\Psi+W\Psi\right). \label{eq:deltaPsi}
\end{equation}
Let us confirm this is a symmetry. Under a general variation, the Lagrangian changes as
\begin{equation}
\delta\mL = \bar{\mathcal{E}} \delta\Psi+ \mathcal{E}\delta\bar\Psi,
\end{equation}
where the equation of motion $\mathcal{E}$ is
\begin{equation}
\mathcal{E} \equiv \Box\Psi + \frac1fr^a\partial_aW\bar\Psi - \frac{W^2+\beta}{f}\Psi.
\end{equation}
In terms of the quantity
\begin{equation}
\bar Q = r^a\partial_a\bar\Psi + W\Psi,
\end{equation}
the variation of the Lagrangian under $\delta\Psi = i\bar Q$ is
\begin{align}
\delta\mL &= i\left(\bar Q\bar{\mathcal{E}}-Q\mathcal{E}\right) \nonumber\\
&= 2\operatorname{Im}Q\mathcal{E}.
\end{align}
Now we calculate $Q\mathcal{E}$ and freely integrate by parts,
\begin{align}
Q\mathcal{E} &= \left(r^a\partial_a\Psi + W\bar\Psi\right)\left(\Box\Psi + \frac1fr^a\partial_aW\bar\Psi - \frac{W^2+\beta}{f}\Psi\right) \nonumber\\
&= W\left(-\partial_a\Psi\partial^a\bar\Psi + \frac{r^a\partial_aW}{f}\left(\Psi^2+\bar\Psi^2\right)-\frac{W^2+\beta}{f}\Psi\bar\Psi\right).
\end{align}
The last line is manifestly real, so that the variation of the Lagrangian vanishes as expected,
\begin{equation}
\delta\mL = 2\operatorname{Im} Q\mathcal E = 0.
\end{equation}
Using similar manipulations we can also calculate the conserved current,
\begin{equation}
J_a = v_a -\bar{v}_a + W\left(\bar\Psi \partial_a\Psi-\Psi\partial_a\bar\Psi\right) - \frac{W^2+\beta}{2f}r_a\left(\Psi^2-\bar\Psi^2\right)
\end{equation}
where we have defined
\begin{equation}
v^a\partial_a = -(\partial_r\Psi)(\partial_t\Psi)\partial_t + \frac{f^2(\partial_r\Psi)^2+(\partial_t\Psi)^2}{2}\partial_r
\end{equation}
such that $r^a\partial_a\Psi\Box\Psi=\nabla_av^a$.

Analogously to the spin-1 case, it is natural to wonder whether this complex master variable is related to the middle-weight Weyl scalar, $\Psi_2$. A new complication in the gravitational case is that $\Psi_2$ has a background value, and accordingly its perturbation $\delta \Psi_2$ is not gauge-invariant. Nevertheless one can construct a gauge-invariant version $\widetilde{\delta\Psi_2}$ which contains the Regge-Wheeler and Zerilli variables \cite{Aksteiner:2010rh},
\begin{equation}
\widetilde{\delta\Psi_2} = \frac{\sqrt{(\ell-1)\ell(\ell+1)(\ell+2)}}{2r^3}\left[\frac{\Lambda}{(\ell+2)(\ell-1)}\Psi_+ + i \Psi_-\right]Y(\theta)
\end{equation}
This is not quite our master variable $\Psi$, as the real (even) piece is a rescaling of the Zerilli variable. We leave a further exploration of this question for future work.

\subsection{Flat-space limit: linearized gravitational duality}

We can gain some physical insight by looking at the flat-space limit, $\rs\to0$. The expression \eqref{eq:psisym} for $\delta\Psi_\pm$ diverges due to the $1/\rs$ scaling in $W(r)$, which can be remedied by sending $\delta \Psi_\pm\to\rs\delta \Psi_\pm$ before taking the limit. In this limit we have an $SO(2)$ symmetry acting on $(\Psi_+,\Psi_-)$ similar to the electromagnetic case,
\begin{subequations}
\begin{align}
\delta\Psi_+ &= -\Psi_-, \\
\delta\Psi_- &= \Psi_+. \label{eq:psisym-flat}
\end{align}
\end{subequations}
Direct calculation shows that, on shell, this duality generates rotations between the Riemann tensor and its dual,
\begin{subequations}
\begin{align}
\delta R_{\mn\ab} &= {\star}R_{\mn\ab},\\
\delta {\star}R_{\mn\ab} &= - R_{\mn\ab},
\end{align}
\end{subequations}
where the dual Riemann tensor is defined as
\begin{equation}
{\star}R_{\mn\ab} = \frac12\epsilon_{\mn\rho\sigma}R^{\rho\sigma}{}_\ab.
\end{equation}
This is the well-known gravitational ``electric-magnetic'' duality, lifted to an off-shell symmetry for linear perturbations around flat space \cite{Henneaux:2004jw}.

We conclude that the symmetry \eqref{eq:psisym} is an extension of electromagnetic duality to Schwarzschild backgrounds. An off-shell duality symmetry has also been found to hold for Minkowski \cite{Henneaux:2004jw}, de Sitter \cite{Julia:2005ze}, and anti-de Sitter backgrounds \cite{Hortner:2019iip}. Adding to this list Schwarzschild, which is less symmetric than the others, raises interesting questions: which other backgrounds possess a linearized duality symmetry, and what physical mechanism underlies these symmetries?

\subsection{Chandrasekhar duality off-shell}

The symmetry \eqref{eq:psisym} can be lifted to a symmetry of the linearized Einstein-Hilbert action in terms of the metric perturbations, \cref{eq:odd-action-22,eq:even-action-22}, analogously to electromagnetism. The calculation itself is cumbersome and not especially enlightening, so we will outline the steps without presenting full expressions. Let us begin with the transformation of the odd-sector variable $h_a$. Using its solution \eqref{eq:ha-sol} and undoing various rescalings, we have
\begin{equation}
\delta h_a = \frac{1}{\sqrt{2\ell(\ell+1)r^2}}\epsilon_\ab \partial^b\left(r\delta\Psi_-\right),
\end{equation}
where $\delta\Psi_-$ is given by \cref{eq:psisym}. That expression is constructed from $\Psi_+$, which we in turn write in terms of even-sector metric perturbations by following the chain of field redefinitions. For the even sector, we vary the expressions in terms of $\Psi_+$ for $h_\ab$ and $\alpha$ or $K$, use \cref{eq:psisym}, and relate $\Psi_-$ to $h_a$ via
\begin{equation}
\Psi_- = \frac{r^3}{2\sqrt{(\ell+2)(\ell-1)}}\epsilon^\ab F_\ab.
\end{equation}
In this way we construct (rather complicated) expressions $\delta h_\mn[h]$ which one can verify by explicit calculation comprise an off-shell symmetry of \cref{eq:odd-action-22,eq:even-action-22}. Interestingly they can be simplified somewhat using the equations of motion, in which case the expressions become entirely local.
A natural question for future investigation is whether the $\delta h_\mn$ constructed this way is equal to a dual potential $\tilde h_\mn$. Since only the electric part of the Weyl tensor has a non-vanishing background value, the linearized duality transformations do not simply rotate $C_{\mn\alpha\beta}$ and $\tilde C_{\mn\alpha\beta}$.

\section{Physical implications: Love numbers}
\label{sec:Love}

Another aspect of black hole perturbation theory in which symmetry has recently been found to play a crucial role is in the computation of \emph{tidal Love numbers}. In particular, the puzzle over the unexpected vanishing of black hole Love numbers \cite{Damour:2009vw,Binnington:2009bb,Porto:2016zng,Chia:2020yla,Charalambous:2021mea} spurred the discovery of underlying symmetry structures \cite{Hui:2021vcv,Hui:2022vbh,Charalambous:2021kcz,Charalambous:2022rre}. It turns out that the duality symmetry which is the focus of this paper also plays a role in the symmetry story for Love numbers.

Consider the Regge-Wheeler action \eqref{eq:odd-action-22-coords} in the static sector, i.e., setting time derivatives to zero,
\begin{align}
\mathcal{L}_\mathrm{odd}^{\omega=0} &= \frac12r^4h_0'^2 + \frac{(\ell+2)(\ell-1)}{2}r^2\left(\frac1fh_0^2-fh_1^2\right), \label{eq:odd-action-static}
\end{align}
where primes denote $r$ derivatives. In the static limit $h_1$ is auxiliary and decouples from $h_0$, so can be consistently set to zero.
The Regge-Wheeler variable $\Psi_-$ is related to $h_0$ by
\begin{equation}
\Psi_- = r^3h_0',\quad h_0 = \frac{f}{r^2}\partial_r\left(r\Psi_-\right).
\end{equation}

In \rcite{Hui:2021vcv} it was shown that the static Regge-Wheeler equation is invariant under \emph{ladder symmetries} which are responsible for the vanishing of tidal Love numbers in the odd sector. These come in the form of raising and lowering operators which relate solutions of the Regge-Wheeler equation to a solution with $\ell$ raised or lowered by one,
\begin{subequations}
\begin{align}
D^+_\ell &= -r^2f \partial_r + \frac{\ell^2+3}{2(\ell+1)}\rs-\ell r,\\
D^-_\ell &= r^2f\partial_r + \frac{\ell^2(\rs-2r)-2\ell(r-\rs)+4\rs}{2\ell}.
\end{align}
\end{subequations}
At the lowest rung of the ladder, $\ell=2$, there is a further symmetry given by $\delta \Psi_-^{\ell=2} = Q_2 \Psi_-^{\ell=2}$, with
\begin{equation}
Q_2 = r^6f\partial_r -3r^5f.
\end{equation}
It follows that any $\ell$ mode is symmetric under the ``horizontal'' ladder symmetry
\begin{equation}
\delta\Psi_-=Q_\ell\Psi_-,
\end{equation}
where $Q_\ell$ is built recursively from $Q_2$,
\begin{equation}
Q_\ell \equiv D^+_{\ell-1}Q_{\ell-1}D_\ell^-.
\end{equation}
Transforming from $\Psi_-$ to $h_0$, we see that the metric transforms under the horizontal odd-sector ladder symmetry as
\begin{equation}
h_0 \to h_0 + \frac{f}{r^2}\partial_r\left[r Q_\ell(r^3h_0')\right], \quad h_1\to h_1.\label{eq:h0sym}
\end{equation}

It is straightforward to check that \cref{eq:h0sym} is a symmetry of \cref{eq:odd-action-static}. However, such a symmetry of the Zerilli equation is not apparent. Indeed, the argument for the vanishing of Love numbers for the Zerilli equation in \rcite{Hui:2021vcv} relied on the fact, as we will show, that the duality invariance \eqref{eq:psisym} implies that the even and odd Love numbers are equal.\footnote{This is to some extent an artifact of our insistence on working with the Regge-Wheeler and Zerilli master equations. The main result in \rcite{Hui:2021vcv} worked with the Teukolsky equation, which, besides not being limited to the Schwarzschild case, contains both even and odd modes.}

Ladder operators for the Zerilli equation can be constructed straightforwardly by sandwiching a Regge-Wheeler ladder operator between two applications of the duality symmetry, e.g., for the horizontal operators,
\begin{equation}
\delta \Psi_{+,\ell} = \left(\partial_{r_\star}-W\right)Q_\ell\left(\partial_{r_\star}+W\right)\Psi_{+,\ell}.
\end{equation}
It would be very interesting to know whether this symmetry is responsible for universal relations such as I-Love-Q \cite{Yagi:2013awa,Yagi:2016bkt}.

\subsection{Equality of Love numbers from gravitational duality}

Let us finish by establishing that the vanishing of the duality Noether current requires the tidal Love numbers in the even and odd sectors to be equal. Following \rcite{Kol:2011vg}, we calculate the Love numbers for static solutions by imposing regularity at the horizon and examining the behavior of the fields at infinity,
\begin{equation}
\Psi_\pm\to \bar\Psi_\pm\left(r^{\ell+1} + \hat\lambda_\pm r^{-\ell}\right), \label{eq:thedefinitionoflove}
\end{equation}
where $\hat\lambda_\pm$ are the Love numbers for the even ($+$) and odd ($-$) sectors and $\bar\Psi_\pm$ are constants.

Since we are looking at static solutions, conservation of the Noether current \eqref{eq:symcons} becomes the statement that the $r_\star$ component \eqref{eq:Jrs} is constant.
First we need to ensure that the duality transformation \eqref{eq:psisym} preserves the boundary conditions, namely, if $\Psi_\pm$ is regular at the horizon, then so is $\partial_{r_\star}\Psi_\pm\mp W(r)\Psi_\pm$. From \cref{eq:superpot} we see that $W(\rs)$ is finite, which leaves us to check that $\partial_{r_\star}\Psi_\pm = f(r)\partial_r\Psi_\pm$ is regular at $r=\rs$. We can see this by solving the Regge-Wheeler and Zerilli equations perturbatively near the horizon,
\begin{equation}
f\partial_r(f\partial_r\Psi_\pm) = V_\pm \Psi_\pm.
\end{equation}
It is convenient to use $f$ as our radial coordinate, so that we can simply expand around $f=0$ to look at the horizon. Using the fact that the Regge-Wheeler and Zerilli potentials both scale as $f$ near the horizon, and that $\partial_r = f'(r)\partial_f \approx \partial_f/\rs$, we have
\begin{equation}
f\partial_f(f\partial_f \Psi_\pm) \approx \frac{(\ell-1)\ell(\ell+1)(\ell+2)\pm3}{\ell(\ell+1)+1} f \Psi_\pm.
\end{equation}
Near the horizon this is solved by
\begin{equation}
\Psi_\pm = c_1^\pm\left(1+\mathcal{O}(f)\right) + c_2^\pm\ln f\left(1+\mathcal{O}(f)\right).
\end{equation}
Regularity at the horizon demands $c_2^+=c_2^-=0$, so that $f\partial_r\Psi_\pm\approx f\partial_f\Psi_\pm\to0$ as $f\to0$.\footnote{We must also check that the subdominant terms do not blow up at the horizon. Assuming that the subdominant terms on the log side go as $f^n \ln f$, then these contribute harmlessly to $f\partial_f\Psi_\pm$ for $n>0$, and moreover they vanish upon perturbatively solving the equation of motion.} So if $\Psi_\pm$ is a solution with boundary conditions suitable for computing Love numbers, then $\tilde\Psi_\pm \equiv \Psi_\pm + \delta\Psi_\pm$ is as well.

Now we simply need to compute $\jrs$ at the horizon and at infinity and equate the two, where for static solutions
\begin{equation}
\jrs = \Psi_+' \Psi_-' + W\left(\Psi_+\Psi_-'-\Psi_-\Psi_+'\right)-(W^2+\beta)\Psi_+\Psi_-.
\end{equation}
We begin by evaluating this at the horizon. Primes denote $r_\star$ derivatives, and we are assuming that $\Psi_\pm$ are regular at the horizon, so $\Psi_\pm' = (1-\rs/r)\partial_r\Psi_\pm=0$ at $r=\rs$. From \cref{eq:superpot} we see $W^2(\rs)+\beta=0$, so that the current vanishes for static solutions with regular boundary conditions,
\begin{equation}
\jrs=0.
\end{equation}
At infinity, we again have $W^2(\infty)+\beta=0$, so the leading-order terms will be those with only one derivative,
\begin{align}
\jrs &\to W(\infty)\left(\Psi_+\Psi_-'-\Psi_-\Psi_+'\right) \nonumber\\
&= -\frac{\bar{\Psi}_+\bar{\Psi}_-}{6\rs} \ell (\ell+1)(\ell-1) (\ell+2)(2\ell+1) \left(\hat\lambda_+-\hat\lambda_-\right).
\end{align}
Since $\jrs=0$ everywhere, we conclude that
\begin{equation}
\boxed{\hat\lambda_+ = \hat\lambda_-,}\label{eq:equallove}
\end{equation}
i.e., the even and odd sectors are forced to have equal Love numbers as a consequence of symmetry.

It turns out that both of these Love numbers are strictly zero \cite{Damour:2009vw,Binnington:2009bb,Porto:2016zng,Chia:2020yla,Charalambous:2021mea}, which is a consequence of a different symmetry than the duality considered in this paper \cite{Hui:2021vcv,Hui:2022vbh,Charalambous:2021kcz,Charalambous:2022rre}, but these conclusions are distinct from each other, i.e., \cref{eq:equallove} does not just say $0=0$. The equality of Love numbers follows from the invariance under duality of the boundary conditions. This is clearly the case for black holes, where regularity at the horizon implies the duality-invariant $c_2^+=c_2^-$, but one could also in principle imagine a horizonless compact object with non-zero but (approximately) equal Love numbers, provided that whatever boundary conditions are chosen at its surface are invariant under duality and that the object is sufficiently compact that $J^{r_\star}\approx0$.

\section{Discussion}
\label{sec:disc}

We have computed the actions for scalar, electromagnetic, and linearized fields on a Schwarszchild background in the $2+2$ formalism. In each case we focused on isolating and canonically normalizing the underlying dynamical degrees of freedom. In the cases of electromagnetism and gravity, this exercise revealed a manifest electric-magnetic duality symmetry, which holds off shell and accordingly can be used to construct conserved quantities.

As a physical application of the Noether current associated to linearized gravitational duality, we showed that duality forces the even- and odd-parity perturbations to have identical tidal responses. Combining this duality with a ``ladder'' symmetry \cite{Hui:2021vcv} which causes the odd Love numbers to vanish therefore extends that particular argument for vanishing Love numbers to even perturbations. It would be interesting to explore whether these symmetries play a role in universal relations for compact objects.

In the case of electromagnetism, we found a clear connection to objects arising in the Newman-Penrose and Geroch-Held-Penrose formalisms: the dynamical master variable is related to the middle-weight Maxwell scalar $\phi_1$. This observation enabled us to derive actions for the Fackerell-Ipser equation and Teukolsky-Starobinsky identities. It would be quite interesting to extend these constructions to the Teukolsky equation for the extreme-weight Maxwell scalars, to gravity, and to Kerr, which is the case of prime astrophysical interest. We leave these questions for future work.

\section*{Acknowledgements}

I am grateful to Lam Hui, Austin Joyce, Riccardo Penco, and Luca Santoni for collaboration, and many insightful discussions, on duality and other topics in black hole perturbation theory.
This work made substantial use of \href{http://xact.es/index.html}{xAct} and the \href{https://people.brandeis.edu/~headrick/Mathematica/}{diffgeo} Mathematica package by Matthew Headrick.
My research is partially supported by funds from the Natural Sciences and Engineering Research Council (NSERC) of Canada. Research at the Perimeter Institute is supported in part by the Government of Canada through NSERC and by the Province of Ontario through MRI.

\appendix

\section{$2+2$ Ricci tensor components}
\label{app:22Ricci}

In this appendix we reproduce the components of the linearized Ricci tensor $\delta R_\ab$ in the $2+2$ split \cite{Sarbach:2001qq,Martel:2005ir,Chaverra:2012bh}, using the partially gauge-fixed metric perturbation \eqref{eq:gauge-h}. We do not present a complete list but focus on components of $\delta R_\ab$ necessary to compute the linearized Einstein-Hilbert action, cf. \cref{eq:action-even,eq:action-odd}.

\subsection{Odd perturbations}

In the odd sector, the only non-zero component is
\begin{align}
\delta R^B_{aA} &= \left(\frac{1}{2r^2}\nabla^b\left(r^4F_\ab\right)-h_a\right)B_A + h_aD^BD_{[A}B_{B]} \nonumber\\
  &= \underbrace{\left(\frac{1}{2r^2}\nabla^b\left(r^4F_\ab\right) + \frac{(\ell+2)(\ell-1)}{2}h_a\right)}_{\delta R^B_a}B_A,
\end{align}
where
\begin{equation}
F_\ab = \partial_ah_b-\partial_bh_a.
\end{equation}
In going to the second line we used the identity
\begin{equation}
D^BD_{[A}B_{B]} = \frac12\ell(\ell+1)B_A.
\end{equation}
To prove this, notice that $D_{[A}B_{B]} \propto \epsilon_\AB$ by symmetry, where the coefficient is
\begin{align}
D_{[A}B_{B]}&=\frac{\epsilon^{CD} D_CB_D}{2}\epsilon_\AB \nonumber\\&= \frac{\ell(\ell+1)}{2}\epsilon_\AB,
\end{align}
where we have used the definition \eqref{eq:BA} of $B_A$. Taking a derivative and using the definition again, the result follows.

\subsection{Even perturbations}

To compute the Lagrangian \eqref{eq:action-even} we need the following components of the perturbed Ricci tensor:
\begin{equation}
\delta R_\ab,\quad \Omega^\AB\delta R_\AB,\quad \delta R^E_a.
\end{equation}
Note that we do not need the piece of $\delta R^E_a$ involving $K$.
The relevant pieces of the relevant components are
\begin{subequations}
\begin{align}
\delta R_\ab & = \left(\frac12r^{-2}\nabla_c\left(r^2\nabla_d\hat h^{cd}\right) - \frac14\Box h\right)g_\ab+\frac2rr_c\hat C^c{}_{\langle\ab\rangle} + \frac1r r_{\langle a}\partial_{b\rangle}h \nonumber\\
&\hphantom{{}=} - r^{-2}\nabla_{(a}\left(r^2\nabla_{b)}K\right) - \frac{\ell(\ell+1)}{r^2}\nabla_{(a}\left(r^2r_{b)}\alpha\right)+\frac12R\hat h_\ab +\frac{\ell(\ell+1)}{2r^2}h_\ab , \\
\Omega^\AB \delta R_\AB &= 2\nabla_a(rr_b\hat h^\ab)  + \frac{\ell(\ell+1)+2}{2}h - r^2\nabla_a\left(r^4\nabla^aK\right) + (\ell+2)(\ell-1)K \nonumber\\
&\hphantom{{}=} - \ell(\ell+1)\left[r^2r^a\partial_a\alpha + (1+3f)r\alpha\right], \\
\delta R^E_a & = \frac12\nabla^b\hat h_\ab-\frac14\nabla_ah+\frac12\partial_a\ln r h - \frac{1}{r^2}\nabla^b\left[r^4r_{[a}\nabla_{b]}\alpha\right] -  r_a\alpha.
\end{align}
\end{subequations}
Angular brackets denote tracefree symmetrization,
\begin{equation}
T_{\langle\ab\rangle} = T_{(ab)}-\frac12T^c{}_cg_\ab.
\end{equation}
In the above we have defined
\begin{align}
\hat C^c{}_\ab = \nabla_{(a}\hat h^c_{b)}-\frac12\nabla^c\hat h_\ab.
\end{align}
The expression for $\delta R_\ab$ contains the term $\nabla_c\hat C^c{}_\ab$ (cf. \rcite{Martel:2005ir}), which we have simplified using the identity
\begin{equation}
\nabla^c\nabla_{(a}p_{b)c}-\frac12\Box p_{ab}-\frac12 g_{ab}\nabla^c\nabla^d p_{cd}=\frac R2p_{ab}
\end{equation}
for symmetric traceless tensors $p_\ab$ in $D=2$,
\begin{equation}
\nabla_c \hat C^c{}_\ab = \frac12R\hat h_\ab + \frac12\nabla_c\nabla_d\hat h^{cd} g_\ab.
\end{equation}

\nocite{apsrev41Control}
\bibliographystyle{apsrev4-1}
\bibliography{biblio}

\end{document}